\documentclass[12pt,a4paper]{article}

\usepackage{amsmath,amssymb}
\usepackage[bookmarks=true]{hyperref}
\usepackage[nosort]{cite}
\usepackage{bbm}

\usepackage{pdfsync}
\makeatletter
\let\old@startsection=\@startsection
\renewcommand{\@startsection}[6]{\old@startsection{#1}{#2}{#3}{#4}{#5}{#6\mathversion{bold}}}
\makeatother

\def\eeq{\end{eqnarray}}

\newcommand{\nn}{\nonumber}
\def\p{\partial}

\def\de{\partial}

\def\=:{=\hspace{-.7em}\raisebox{1.1ex}{.}\hspace{.1em}\raisebox{-0.2ex}{.} }

\newcommand{\beqn}{\begin{eqnarray}}
\newcommand{\eeqn}{\end{eqnarray}}
\newcommand {\beq}{\begin{eqnarray}}
\def\eeq{\end{eqnarray}}
\newcommand {\eeqq}{\end{eqnarray}}
\newcommand {\non}{\nonumber\\}
\def\nn{\nonumber\\}

\newcommand {\del}{\partial}

\newcommand {\brc}{\langle}
\newcommand {\ckt}{\rangle}

\newcommand {\Tr}{{\rm Tr}\,}

\newcommand{\cp}{\mathbf{CP}^1}

\makeatletter \@addtoreset{equation}{section} \makeatother

\makeatletter
\let\old@makecaption=\@makecaption
\def\@makecaption{\small\old@makecaption}
\makeatother

\makeatletter
\def\mr@ignsp#1 {\ifx\:#1\@empty\else #1\expandafter\mr@ignsp\fi}%
\newcommand{\multiref}[1]{\begingroup
\xdef\mr@no@sparg{\expandafter\mr@ignsp#1 \: }%
\def\mr@comma{}%
\@for\mr@refs:=\mr@no@sparg\do{\mr@comma\def\mr@comma{,}\ref{\mr@refs}}%
\endgroup}
\makeatother

\ifx\href\asklfhas\newcommand{\href}[2]{#2}\fi

\begin{document}

\begin{flushright}
IFUP-TH/2013-18, OCU-PHYS 392
\end{flushright}
\vspace{0.4 cm} 

\renewcommand{\thefootnote}{\arabic{footnote}}
\setcounter{footnote}{0}
\begin{center}%
{\Large\textbf{\mathversion{bold}
 Non-Abelian Vortices with an Aharonov-Bohm Effect}
\par}

\vspace{0.7cm}%

\textsc{ Jarah Evslin$^{1,2}$, Kenichi Konishi$^{3,4}$,  Muneto Nitta$^{5}$, \\ Keisuke Ohashi$^{6}$,  Walter Vinci$^{7}$}

\vspace{10mm}

{\small 

\textit{$^{1}$ TPCSF, IHEP, Chinese Acad. of Sciences, Beijing, China  }

\textit{$^{2}$  Theoretical physics division, IHEP, Chinese Acad. of Sciences, Beijing, China  }

\textit{$^{3}$  Department of Physics, ``Enrico Fermi`'',    University of Pisa,  \\
Largo Pontecorvo,3, 56127, Pisa, Italy }
\\
\textit{$^{4}$  INFN, Sezione di Pisa,  
Largo Pontecorvo,3, 56127, Pisa, Italy}
\\
\textit{$^{5}$  Department of Physics, and Research and Education Center 
for Natural Sciences,\\ Keio University, 4-1-1 Hiyoshi, Yokohama, 
Kanagawa 223-8521, Japan}
\\
\textit{$^{6}$ Department of Physics, Osaka City University, Osaka, Japan} \\
\textit{$^{7}$ London Centre for Nanotechnology and Computer Science, 
University College London, 17-19 Gordon Street,
London,  WC1H 0AH, United Kingdom.
}

} 

\vspace{.6cm}

\thispagestyle{empty}
\texttt{jarah(at)ihep.ac.cn}, 
\texttt{konishi(at)df.unipi.it}\\
\texttt{nitta(at)phys-h.keio.ac.jp}, 
\texttt{ohashi(at)sci.osaka-cu.ac.jp}\\
\texttt{w.vinci(at)ucl.ac.uk}

\par\vspace{0.2cm}

\vfill
\textbf{Abstract}\vspace{5mm}

\begin{minipage}{12.7cm}

 The interplay of gauge dynamics and flavor symmetries often leads to
remarkably subtle phenomena in the presence of soliton configurations. Non-Abelian vortices -- vortex solutions with continuous internal orientational moduli -- provide an example.  Here we study the effect of weakly gauging a $U(1)_{R}$ subgroup of the flavor symmetry on such BPS vortex solutions. Our prototypical setting consists of an $SU(2)\times U(1)$ gauge theory with $N_{f}=2$ sets of fundamental scalars that break the gauge symmetry to an ``electromagnetic"  $U(1)$.   The weak $U(1)_{R}$ gauging  converts the well-known $\cp$ orientation modulus $|B|$ of the non-Abelian vortex into a parameter characterizing the strength of the magnetic field that is responsible for the Aharonov-Bohm effect.  As the phase of $B$ remains a genuine zero mode while the electromagnetic gauge symmetry is Higgsed in the interior of the vortex, these solutions are superconducting strings.

\end{minipage}

\vspace{3mm}

\vspace*{\fill}

\end{center}

\newpage

\section{Introduction}

Topological solitons play a fundamental role in various physical systems, especially in those characterized by gauge interactions. The interplay of the classical or quantum gauge dynamics with the global symmetries in such systems  leads to remarkably rich and subtle phenomena.  Of particular interest among these is a class of vortex solutions carrying continuous orientational zero modes: non-Abelian vortices~\cite{Hanany:2003hp}~--~\cite{Shifman:2009zz}. 

These typically arise when the (complete) gauge symmetry breaking supports vortex solutions, and at the same time the vacuum is invariant under a color-flavor diagonal group so that the system is in the so-called color-flavor locked phase~\cite{Alford:2007xm}. As the individual vortex breaks the exact global symmetry, it develops orientational zero modes which can fluctuate on the vortex world sheet.  At low energies these fluctuations can be described by an appropriate $2D$ sigma model that has its own nontrivial, large-distance, quantum dynamics. 
A longstanding goal concerning non-Abelian vortices is to find solutions describing a non-Abelian vortex-monopole complex~\cite{Auzzi:2003fs},~\cite{Auzzi:2003em},~\cite{Konishi70}, \cite{Nitta:2010nd} -- \cite{Eto:2011cv}. 

Here we examine yet another, little studied, question regarding non-Abelian vortex systems:  what happens if a part or the whole of the exact global (color-flavor diagonal) symmetry is weakly gauged?   In Ref.~\cite{KNV}, we have initiated the study of such effects by weakly gauging the entire exact flavor symmetry of these vortex solutions. In a sense this was a simpler question: the answer is that now any color-flavor rotation of a given solution is a genuine global gauge transformation, rendering all charge one vortex solutions gauge equivalent whatever their orientations. Stated differently, a mini-Higgs mechanism is at work in the vortex worldsheet, transforming the orientational zero modes into light, propagating modes. As a result of the non-Abelian nature of the gauge groups involved, these light modes are unstable and     
decay into massless $4D$ gauge bosons. 

Here we turn our attention to the case where only a part of the global symmetry group is gauged. As a simple concrete model  we take an $SU(2)_{L} \times U(1)$ gauge theory with two complex scalars in the fundamental representation, in which a $U(1)_{R}\subset SU(2)_{R}$ subgroup of the flavor symmetry is weakly gauged\footnote{Here $R$ means ``acting from the right'', i.e, on the flavor indices and is unrelated to the $R$ symmetry of supersymmetric theories.}.  The $SU(2)_{L} \times U(1)$ and $U(1)_{R}$ gauge symmetries are broken by the scalar VEVs, but  a $U(1)^{em}$ subgroup remains unbroken.  These $4D$ massless gauge modes interact with the 
$2D$ zero modes on the vortex worldsheet.  The result is rather unexpected and elegant. We find that the two-dimensional vortex moduli (partially) survive the $U(1)_{R}$ gauging,  but the complex parameter $B$ inherited from the original $\cp$  vortex moduli acquires a new physical meaning.  The modulus $|B|$ can still be interpreted as a truly $2D$ vortex modulus which  now characterizes the  magnetic flux carried by the vortex  responsible for an  Aharonov-Bohm effect
a particle with unit  $U(1)_{R}$ charge wil experience
 \cite{Aharonov:1959fk}  in 
going around the vortex far from the core. 
The phase Arg$(B)$, on the other hand, is eaten by the $4D$ gauge bosons.
An AB effect on vortices have been studied earlier \cite{Alford:1988sj}. Compared to that case, 
a peculiar feature of our case is that 
the AB phase depends on the modulus 
$|B|$. 

The question of $U(1)$ gauging was originally raised 
in the context of dense quark matter.
QCD in the high baryon density limit is 
believed to be in the color-flavor locked phase 
with  diquark condensation~\cite{Alford:2007xm}, 
where the $SU(3)$ color symmetry and 
the $SU(3)$ flavor symmetry of the three light quarks 
are spontaneously broken to the diagonal subgroup.
In this context, non-Abelian vortices appear~\cite{Balachandran:2005ev} -- \cite{Nakano:2007dr}
which differ from local non-Abelian vortices 
for which the $U(1)$ group is a global symmetry. 
Nevertheless, they possess normalizable 
orientational zero modes~\cite{Eto:2009bh}, \cite{Eto:2009tr}.
Taking into account the $U(1)$ 
electromagnetic coupling corresponds 
to gauging a $U(1)$ subgroup of the 
flavor $SU(3)$ symmetry,  
some consequences of which were 
studied in Refs.~\cite{Hirono:2012ki} -- \cite{Vinci:2012mc}.
See Ref.~\cite{Eto:2013hoa} for a recent review.

\subsection{Orientational $\cp$ modes of the standard non-Abelian vortex \label{standard}}
 
Before introducing the weak gauging of a part of the flavor symmetry,  let us first briefly review a few salient  features of the non-Abelian vortex. The simplest example is an $SU(2)\times U(1)$ gauge theory with two scalar fields transforming in the fundamental representation, $Q = \left(\begin{array}{cc}q^1 & q^2\end{array}\right)$, written in a color-flavor mixed $2 \times 2$ matrix form.   The action is
\begin{eqnarray}
S & = & \int d^{4}x \left\{   \frac1{4 g^{2}}(F^{0}_{\mu\nu})^{2}+\frac1{4 g^{2}}(F^{a}_{\mu\nu})^{2}+ |\nabla_{\mu} q^{A}|^{2} + \frac{g^{2}}{8}\left(\bar q^{A}\tau^{a}q^{A}  \right)^{2 }+\frac{g^{2}}8(|q^{A}|^{2}-2\xi)^{2} \right\} \,,\nonumber \\
  \label{eq:kinkact}
\end{eqnarray}
where
\begin{equation}
\nabla_{\mu}q^{A}=\partial_{\mu}q^{A} +  \frac{i}{2}  A^{0}_{\mu}q^{A} + i \frac{\tau^{a}}2 A^{a}_{\mu}q^{A}\,,  \qquad A=1,2\;.
\end{equation}
As is often done in the recent literature, the matter content and the potential terms are chosen such that the model above can be extended to have an $\mathcal N=2$ supersymmetry.   As a consequence,  the scalar quartic couplings are set to the critical value so that the classical equations for the soliton configurations become first order differential equations, in the ``self-adjoint'', or BPS,  form. 

In the presence of a nonzero parameter $\xi$,  the system is in a Higgs vacuum:
\beq
Q_{vev}= q_{i}^{A}=
\left(
\begin{array}{cc}
 \sqrt\xi & 0    \\
  0 &   \sqrt\xi
\end{array}
\right)\,.
\label{vacuum}  \eeq
The gauge and flavor symmetries are completely broken, but there remains an unbroken color-flavor diagonal  $SU(2)_{C+F}$  global symmetry  (i.e., it is in color-flavor locked phase).
As 
\beq  \pi_{1}\left({SU(2)\times U(1)\over {\bf Z}_2}\right)=  {\mathbbm Z}\;, 
\eeq
the system possesses stable, nonsingular vortices.  An individual vortex solution breaks the $SU(2)_{C+F}$ global symmetry to  a $U(1)$ subgroup and so it develops an orientational modulus $B\in\cp=SU(2)/U(1)$.  Indeed,
the vortex solution  with a generic orientation and in the regular gauge takes the form
\begin{align}
Q   &= U
\begin{pmatrix}
e^{i \varphi}  \phi_1(r) & 0 \\
0 & \phi_2(r)
\end{pmatrix} U^{-1}
= \frac{e^{i \varphi} \phi_1(r)+\phi_2(r)}{2}\,  \mathbf{1}_{2}
+ \frac{ e^{i \varphi} \phi_1(r)-\phi_2(r)}{2} \,U T U^{-1} \ , \non
A_i &= -\frac{1}{2}\epsilon_{ij}\frac{x^j}{r^2} \left[
(f(r)-1) \, \mathbf{1}_{2} + (f_{\rm NA}(r)-1) \, U T U^{-1} \right] \ ,
\qquad  i=1,2 \   \label{genericorient}
\end{align}
\beq  T = {\rm diag}\,\left(1, -1\right) = \tau^{3},  \label{Tmatrix} \eeq
where the boundary conditions are
 \begin{align}
\phi_{1,2}(\infty) = \sqrt{\xi} \ , \qquad
\phi_1(0) = 0 \ , \quad
\p_r\phi_2(0) = 0 \ , \quad
\label{eq:bcs1}
\end{align}
 \begin{align}
f(\infty)=f_{\rm NA}(\infty) = 0 \ , \qquad
f(0) = f_{\rm NA}(0) = 1 \ .
\label{eq:bcs2}
\end{align}
The ``reducing matrix''  $U$ has the form 
\begin{align}
U =
\begin{pmatrix}
\mathbf{1} & - B^\dag \\
0 & \mathbf{1}
\end{pmatrix}
\begin{pmatrix}
X^{\frac{1}{2}} & 0 \\
0 & Y^{-\frac{1}{2}}
\end{pmatrix}
\begin{pmatrix}
\mathbf{1} & 0 \\
B & \mathbf{1}
\end{pmatrix}
=
\begin{pmatrix}
X^{-\frac{1}{2}} & - B^\dag Y^{-\frac{1}{2}} \\
B X^{-\frac{1}{2}} & Y^{-\frac{1}{2}}
\end{pmatrix}
\label{eq:Umatrix} \ ,
\end{align}
with the matrices $X$ and $Y$ defined by
\beq
X\equiv\mathbf{1} + B^\dag B \ , \quad
Y\equiv\mathbf{1}  + B B^\dag \; .   \label{X&Y}
\eeq
  The vortex tension, 
\[    T = 2 \pi \xi
\]
does not depend on  the $\cp$ coordinate  $B$.

More generally, perturbations of these solutions can be described by promoting the modulus $B$ to a collective coordinate which depends upon the worldsheet coordinates of the vortex.  The fluctuations of $B$ are then described by a worldsheet   $\cp$ sigma model: 
\beq   S_{eff}=   \frac{4\pi}{g_{L}^2}   \int dtdz \;   \frac{1}{(1 + |B|^{2})^{2}}  \del_{\alpha} B^{*} \del^{\alpha} B\;.   \label{CP1sigma}
\eeq
Our main interest below  is to determine the fate of these $\cp$ collective coordinates in the presence of an external, weak $U(1)_{R}$ gauge field.

\section{The model, BPS equations and vortex solutions}

\subsection{The model and BPS vortex equations}
The model that  we consider in this paper is the same  $SU(2)_{L}\times U(1)_{0}$ gauge theory with $N_{f}=2$ 
flavors of scalar fields as was described above, but we weakly gauge 
   a $U(1)_{R} \subset SU(2)_{R}$  subgroup of the flavor symmetry. 
 The action is then
\begin{eqnarray}
S    & = & \int d^{4}x \left\{   \frac1{4 }(F^{0}_{\mu\nu})^{2}+\frac1{4 }(F^{a}_{\mu\nu})^{2}+ |\nabla_{\mu} q^{A}|^{2} + \frac{g_{L}^{2}}{8}\left(\bar q^{A}\tau^{a}q^{A}  \right)^{2 }+\frac{g_{0}^{2}}8(|q^{A}|^{2}-2\xi)^{2} \right\}+ \nonumber \\
& +& \frac{1}{4}(F^{R\, 3}_{ \mu\nu})^{2}+ \frac{g_{R}^{2}}{8}\left(q^{A}\tau^{R\, 3}\bar q^{A}  \right)^{2 }=\nonumber \\
&= &  \int d^{4}x \, \Tr\left\{   \frac1{2 }F_{\mu\nu}^{2}+ |\nabla_{\mu} Q|^{2} +\frac{g^{2}}4(\bar Q  Q -\xi)^{2} + \frac1{2}(F^{R}_{\mu\nu})^{2}\right\}+\frac{g_{R}^{2}}8   (\Tr (\bar Q  Q \tau^{ 3}))^{2}\,,\nonumber \\
  \label{U(1)gauging}
\end{eqnarray}
where 
\begin{eqnarray}
&F_{\mu\nu}=F_{\mu\nu}^{a}\frac{\tau^{a}}2\;, \quad F_{\mu\nu}^{R}=F_{\mu\nu}^{R\, 3}\frac{\tau^{3}}2\;, \quad A_{\mu}=A^{a}_{\mu}\frac{\tau^{a}}2\;, \quad A_{\mu}^{R}=A_{\mu}^{R}\frac{\tau^{ 3}}2\;;  & \nonumber \\
\nonumber \\
&  \nabla_{\mu}Q=\partial_{\mu}Q + i g_{0} A_{\mu}^{(0)}  Q +  i g_{L} A_{\mu} Q +  i g_{R} Q A^{R}_{\mu}\,.&
\end{eqnarray}
The vacuum  is  the same as in Eq.~(\ref{vacuum}), 
\beq
\brc Q  \ckt= \brc q_{i}^{A} \ckt= 
\left(
\begin{array}{cc}
 \sqrt\xi & 0    \\
  0 &   \sqrt\xi
\end{array}
\right)\,.
\label{vacuumbis}  \eeq
The $U(1)_0$, $SU(2)_{L}$ and $U(1)_{R}$ gauge groups are all  broken, but a combination  
of  $U_{\tau^{3}}(1)\subset SU(2)_{L}$ and $U(1)_{R}$ remains unbroken.  A gauge boson
\beq   A_{\mu}^{em}=   \frac{g_L}{\sqrt{g_{L}^{2}+ g_{R}^{2}}} A_{\mu}^{R} - \frac{g_R}{\sqrt{g_{L}^{2}+ g_{R}^{2}}}   A_{\mu}^{L\, 3},  
\label{em}   \eeq
remains massless in the bulk: we call it  ``electromagnetic'' in analogy with the situation in the  Weinberg-Salam theory.\footnote{As we need several coupling constants here is the summary.  The unbroken electromagnetic field is coupled with the coupling
\[   e \equiv  \frac{g_{R} g_{L}}{\sqrt{g_{L}^{2}+ g_{R}^{2}}};\]  while the broken gauge field $B_{\mu}$ has coupling $g' =  \sqrt{g_{L}^{2}+ g_{R}^{2}}.$   Throughout, we take  $g_{0}$ and $g_{L}$ to be of the same order of magnitude, whereas $g_{R} \ll g_{L}$ hence   $e \ll g_{L}, g_{0}$.  }  
All other gauge bosons, the orthogonal combination
\beq  B_{\mu} = \frac{g_L}{\sqrt{g_{L}^{2}+ g_{R}^{2}}}   A_{\mu}^{L \, 3} +   \frac{g_R}{\sqrt{g_{L}^{2}+ g_{R}^{2}}}   A_{\mu}^{R}\;, \label{broken}
\eeq
$A_{\mu}^{L\, \pm}$, and   $A^{(0)}_{\mu}$ via the Higgs mechanism acquire a mass of order of $M \sim g_{L} \sqrt{\xi} \sim g_{0} \sqrt{\xi}$. 
The inverse of  (\ref{em}) and (\ref{broken}) is
\beq    A_{\mu}^{R} = \frac{g_L}{\sqrt{g_{L}^{2}+ g_{R}^{2}}}   A_{\mu}^{em} +   \frac{g_R}{\sqrt{g_{L}^{2}+ g_{R}^{2}}}   B_{\mu}\;, \qquad 
A_{\mu}^{L\, 3} = \frac{g_L}{\sqrt{g_{L}^{2}+ g_{R}^{2}}}   B_{\mu} -  \frac{g_R}{\sqrt{g_{L}^{2}+ g_{R}^{2}}}   A_{\mu}^{em}\;.\label{inverse}
\eeq

Ordinarily, this would be the end of the story:  $B_{\mu}, A_{\mu}^{L\, \pm}, A^{(0)}_{\mu}$ are  massive fields and cannot propagate farther than $1/M$;  the long-distance physics is dominated by the massless photon  $A_{\mu}^{em}$. However in the presence of a vortex, a one dimensional infinitely long ``hole'' in the condensate,   the situation is a little subtler. Of course, one knows from the standard ANO vortex that a ``massive'' gauge boson  can become massless along the vortex core, the associated magnetic field penetrating  the vortex core  and giving rise to important physical effects such as the vortex tension, a magnetic flux, vortex interactions, etc. 
In the case of a non-Abelian vortex these features are also present, because  in the simplest context of a $U(N)$ theory the latter may be considered to be an ANO vortex embedded in a corner of the color-flavor mixed space. The interesting phenomena related to the internal orientational zero modes and their dynamics  arise from the fluctuation of
this embedding direction. 

In the present case, where $U(1)_{R}$ weak external gauge interactions break the color-flavor rotational symmetry, yet one more, perhaps less familiar,  effect arises.  In a vortex solution of a generic orientation, some combination of scalar fields $Q$  vanish along the vortex core, meaning that some of the ``massive'' gauge bosons become massless.  Part of the effect goes, as in the ANO vortex, into producing a magnetic flux and the associated vortex energy (tension).  On the other hand, far from the core another  combination of the gauge fields, the electromagnetic field, becomes massless  allowing a nontrivial  Wilson loop at infinity.  In our setting we will find that the value of this Wilson loop is {\it unrelated} to the vortex tension.  As there are no electrically charged condensates at infinity,  the electromagnetic Wilson loop can have any value and so it is observable via an Aharonov-Bohm (AB) effect: electrically charged particles circumnavigating the vortex acquire an AB  phase. 

Which components of the gauge fields become massless at the vortex core depends on the particular solution considered,
 and it might appear that it is quite a complicated matter to disentangle these two effects.  Fortunately the BPS nature of our systems allows us to determine in detail the solutions, with the asymptotic behavior of all gauge fields explicitly exhibited.  Knowing them,  and if we are interested only in the observable effects far from the vortex,  we shall be able to describe unambiguously the new AB effect associated with the weak $U(1)_{R}$ gauge field and the massless $A^{em}_{\mu}$.

In obtaining the minimum tension vortex solutions, we will take advantage of the basic fact  is that it is possible to write a BPS completion, even in the presence of the $U(1)_{R}$ gauge field, for static vortex configurations: 
\begin{eqnarray}
S & = & \int d^{2}x\, \Tr\left\{  \left( F_{12}+ \frac{g_{L}}2(Q \bar Q -\xi)\right)^{2} +  \left( F^{R}_{12}+ \frac{g_{R}}2\tau^{R\, 3} \Tr (\bar Q Q\, \frac{\tau^{R\, 3}}{2})  \right)^{2}+\right. \nonumber \\
 & + & |\nabla_{1} Q+i \nabla_{2} Q|^{2} + g_{L}  \, \xi\,F_{12}
  -  \left.\frac12\epsilon_{ij}\partial_{i}\left(i \nabla_{j}Q\bar Q- Q\,i\nabla_{j}\bar Q \right) \right\}
 \label{BPScompletion}
\end{eqnarray}
where the tension depends only on the winding of $U(1)_{0}$.
The BPS equations read 
\beq&& F_{12}=- \frac{g_{L}}{2} (Q \bar Q -\xi)  \;; \label{bps1}  \\
 &&F^{R}_{12} =  - \frac{g_{R}}2 \tau^{R3} \Tr (\bar Q Q\, \tau^{R3}/2)  \;; \label{bps2}  \\
 &&  {\bar D} Q =   \nabla_{1} Q+i \nabla_{2} Q  =  0\;.  \label{bps3}
\eeq

\subsection{BPS solution with $B=0$ \label{BPSB0}}

Several BPS solutions can be found by inspection. 
A vortex solution in which the scalar fields take the color-flavor diagonal form, 
\beq   Q =    \left(\begin{array}{cc}e^{i \varphi} \phi_1(r) & 0 \\0 & \phi_2(r)\end{array}\right)\;,
\eeq
whereas  $A^{\pm}_{i} \equiv 0$  (as in the ``$B=0$" solution of the standard non-Abelian vortex, Eq.~(\ref{genericorient})), can be found easily.  The profile functions satisfy the boundary conditions, Eq.~(\ref{eq:bcs1}).
The solution takes the form
 \beq     A_{R\, i}(x) = \frac{g_{R}}{   \sqrt{g_{R}^{2}+  g_{L}^{2} }}  \, B_{i}(x)\;; \qquad  A_{L\,  i}^{3}(x) =\frac{g_{L}}{   \sqrt{g_{R}^{2}+  g_{L}^{2} }}   \, B_{i}(x)\;. \label{B=0sol1} 
\eeq
 In other words,   Eq.~(\ref{bps2}) and the non-Abelian part  of  Eq.~(\ref{bps1}) are identical, whereas only the combination 
 \beq    g_{L} A_{L}^{3} +  g_{R} A_{R}  =  g^{\prime}\, B_{i}  \label{B=0sol2}
 \eeq
 enters the third BPS equation.    The Abelian  $U(1)_{0}$ field $A^{(0)}_{\mu}$ is as in Eq.~(\ref{genericorient}).

In order to have a finite energy configuration the kinetic term 
\[ \nabla_{i} Q  =   (\de_{i} +i g_{0}A_{\mu}^{(0)} + i  g^{'} B_{i} ) Q\;,  \qquad g^{\prime}= \sqrt{g_{L}^{2}+ g_{R}^{2}}    \]  
 must approach zero asymptotically. This means that  non-vanishing gauge fields (Eq.~(\ref{broken})) must, {\it  in the regular gauge}, approach 
 \beq     g_{0}  A_{i}^{(0)} \to  -\frac{1}{2}\epsilon_{ij}\frac{x^j}{r^2} 
 \left(\begin{array}{cc}1 & 0 \\0 & 1 \end{array}\right)\;;
  \eeq
 \beq     g^{'}  B_{i} =  g_{L}\,  A_{i}^{L \, 3 } +  g_{R}\,  A_{i}^{R} \to  -\frac{1}{2}\epsilon_{ij}\frac{x^j}{r^2} 
 \left(\begin{array}{cc}1 & 0 \\0 & -1 \end{array}\right). \label{asympto}
  \eeq
In fact the condition is even stronger: $A_{i}^{L \, 3 }$ and  $A_{i}^{R}$ are proportional,   $A_{i}^{L \, 3 }/g_{L} = A_{i}^{R}/g_{R}$  (Eq.~(\ref{B=0sol1})).
Combining this  and Eq.~(\ref{asympto}), one finds that  at large $r$ 
\beq     A_{i}^{R}  \to  \frac{g_{R}}{g_{L}^{2} + g_{R}^{2}}   (-\frac{1}{2}) \epsilon_{ij}\frac{x^j}{r^2} \left(\begin{array}{cc}1 & 0 \\0 & -1 \end{array}\right),
\quad   A_{i}^{L\, 3}  \to  \frac{g_{L}}{g_{L}^{2} + g_{R}^{2}}   (-\frac{1}{2}) \epsilon_{ij}\frac{x^j}{r^2} \left(\begin{array}{cc}1 & 0 \\0 & -1 \end{array}\right)\;.
\label{B=0sol3}  \eeq
When a probe particle carrying a unit $U(1)_{R}$ charge $+1$  follows a large circle  around the vortex, it will acquire  an AB phase equal to
\beq   g_{R}  \oint dx^{i}  \,  A_{i}^{R} = g_{R}\int d^{2}x\,   F_{12}^{R}  =  -   2\pi  \frac{g_{R}^{2}}{g_{L}^{2} + g_{R}^{2}} \simeq  -    \frac{2 \pi g_{R}^{2}}{g_{L}^{2}} \;. \label{eq:AB1}
\eeq

\subsection{$|B|= 1$ solution \label{BPSB1}}

 The solution with $|B|=1$ can also be found easily.  Setting
 \beq   B =  e^{i \delta}, 
 \eeq
 in Eq.~(\ref{genericorient})-Eq.~(\ref{eq:Umatrix}), one finds the scalar field configuration 
 \beq   Q &=& U\left(
\begin{array}{cc}
 e^{  i \,  \varphi}   \phi_1(r) & 0  \\
  0 &  \phi_2(r) \\
  \end{array}\right)U^{-1}=   \frac{1}{\sqrt{2}}      \left(\begin{array}{cc}  e^{i\varphi} \phi_1+\phi_2 &  e^{-i \delta} (e^{i\varphi} \phi_1-\phi_2) \\ e^{i \delta} (e^{i\varphi} \phi_1-\phi_2)  & e^{i\varphi} \phi_1+ \phi_2\end{array}\right)
 \nonumber \\
&=& \frac{e^{i\varphi} \phi_{1}+\phi_{2}}{2}  {\mathbbm 1} +  \frac{e^{i\varphi} \phi_{1}-\phi_{2}}{2} (\tau^{1}  \cos \delta  +  \tau^{2}  \sin \delta)\;. \label{B=1sol}
 \eeq
 As  ${\bar Q} Q$   is orthogonal to the direction $\tau^{3}$ of the right $U(1)_{R}$,
 \[   \Tr \, {\bar Q} Q  \,  \tau_{R}^{3}=0\;, 
 \]
  one sees from Eq.~(\ref{BPScompletion})  that  a BPS solution can be constructed by setting  $A_{R\,i}\equiv 0$ and by choosing 
  $A^{L}_{i}= A_{i}$ and $A^{(0)}_{i}$ as in  (\ref{genericorient}): 
 \beq   g_{L} {A}^{L}_{i}(x) = -\frac12\,(\tau^{1}  \cos \delta  +  \tau^{2}  \sin \delta)  \, \epsilon_{ij}\,\frac{x_j}{r^2}\,
[1-f_3(r)]\;, \eeq
\beq   A^{(0)}_{i}(x) = - \epsilon_{ij}\,\frac{x_j}{r^2}\,
[1-f(r)]\;.
\eeq
As  $ A^{R}_{i}(x) \equiv 0,$  $\forall x$, there is no AB effect associated with the $U(1)_{R}$ gauge interactions.

 \subsection{BPS solution with $B= \infty$ \label{BPSB2}}
 
 Using Eq.~(\ref{genericorient}) and Eq.~(\ref{eq:Umatrix}) one finds that  the $B=\infty$ vortex has a squark condensate of the form, 
    \beq   Q =    \left(\begin{array}{cc}   \phi_2(r)  & 0 \\ 0 & e^{i \varphi} \phi_1(r) \end{array}\right)    \stackrel{r \to \infty}   {  \longrightarrow }\sqrt{
   \xi}    \left(\begin{array}{cc}   1  & 0 \\ 0 & e^{i \varphi}  \end{array}\right)\;.    \label{Binfvol}  \eeq
In the model without $U(1)_{R}$ weak gauge interactions this corresponds to the origin of the second patch of the  $\cp$ vortex moduli space, the south pole. 
The vortex equations are symmetric under an exchange of the two flavors combined with a reflection of the moduli space $\cp$, which maps $B\longrightarrow 1/B$.  Therefore this solution is essentially identical to the $B=0$ vortex with the two flavors are interchanged.  As a result the BPS solution can be constructed, in the presence of the $U(1)_{R}$ weak gauge fields,  as in Subsection~\ref{BPSB0}, taking into account certain sign changes (i.e.,  in Eqs.~(\ref{asympto}) and (\ref{B=0sol3})). A probe particle carrying a unit $U(1)_{R}$ charge $+1$  going around the vortex will this time obtain an AB phase of
\beq    g_{R} \oint dx^{i}  \,  A_{i}^{R} =  g_{R}\int d^{2}x\,   F_{12}^{R}  =  2\pi  \frac{g_{R}^{2}}{g_{L}^{2} + g_{R}^{2}}  \simeq     \frac{2 \pi g_{R}^{2}}{g_{L}^{2}}  \;, 
\eeq
i.e.,  of the same magnitude as the $B=0$ vortex but with the opposite sign.

\subsection{A general BPS solution \label{BPSB3}}

To show the existence of other BPS solutions interpolating between the $B=0$, $B=1$ and $B=\infty$ solutions found above requires a more careful analysis, to be presented in the next section. Nevertheless,   
it is easy to obtain the $U(1)_{R}$ flux of such a general solution if one {\it assumes} that it exists and is given by the deformation in $g_{R}\ne 0$ of the unperturbed BPS vortices  Eq.~(\ref{genericorient}) -- Eq.~(\ref{X&Y}) characterized by the $\cp$ coordinate $B$. The point is that to find the BPS solution to first order in $g_{R}$  it is sufficient (see Eq.~(\ref{em})) to know    $A^{L\,3}_{\mu}$  to the zeroth order and $A^{R}_{\mu}$ to first order.  The former is given by Eq.~(\ref{genericorient}) -- Eq.~(\ref{X&Y}), whereas the latter is given by the BPS equation (\ref{bps2}) in which the right hand side is replaced by the scalar fields in the zeroth-order solution Eq.~(\ref{genericorient}). One finds indeed
\beq  F^{R}_{12} \simeq   -    {g_{R}} \Tr (\bar Q Q\, \tau^{R3}/2)|_{g_{R}=0} =  -    \frac{g_{R}}{2}   
 ( \phi_{1}^{2}- \phi_{2}^{2} ) \, \frac{1 - |B|^{2}}{1 + |B|^{2}}\;, 
  \label{bps2Bis} \eeq 
but by virtue of a zeroth order BPS equation (see for instance, Eq.~(3.8) of Ref.~\cite{Auzzi:2003fs}, 
or   Eq.~(16)  of Ref.~\cite{GJK})  
\beq    \frac{1}{r}   \partial_r  f_{\rm NA} = \frac{g_{L}^{2}}{2}
   (\phi_1^2 - \phi_2^2  )\;, \qquad f_{\rm NA}(0)=1, \quad   f_{\rm NA}(\infty)=0\;.   \eeq 
The  $F^{R}_{12}$ flux  (times $g_{R}$)   is given by 
 \beq
g_{R}\int  d^{2}x \, F^{R}_{12} =  \frac{g_{R}^{2}}{g_{L}^{2}}   \frac{1 - |B|^{2}}{1 + |B|^{2}}  \,  2\pi  \int_{0}^{\infty}  dr \, r  \,  \frac{1}{r}   \partial_r  f_{\rm NA} =  
   \frac{2 \pi g_{R}^{2}}{g_{L}^{2}}   \frac{ |B|^{2}-1}{ |B|^{2}+1}\;, \quad 0 \le |B| \le \infty     \label{simple}
 \eeq
 (where Eq.~(\ref{eq:bcs2}) was used).
 This correctly reproduces the results for the $B=0,1,$ and $ \infty$ solutions found above. 
 It is clear from this derivation that the AB effect in Eq.~(\ref{simple}) is gauge invariant, both with respect to the $SU(2)_{L}\times U(1)_0$ and to the weak 
 $U(1)_{R}$ gauge groups.

  \section{Moduli matrix and  the master equations}
 
Showing that there are indeed generic BPS vortex solutions interpolating between those with $B=0$, $B=1$ and $B= \infty$ is somewhat more difficult 
as the color-flavor rotations are no longer exact symmetry operations.  Below we will instead appeal to the powerful moduli-matrix method
developed in  Ref.~\cite{Isozumi:2004vg,Eto:2005yh,Eto:2006pg}.  Before beginning the analysis, let us note that the index for the dimension of the BPS vortex moduli space  
\beq 
\mathcal{I} =  N\, N_{f}\,   \nu\;,
\eeq
(i.e., the number of the zero modes)~\cite{Hanany:2003hp},~\cite{Eto:2009bg} does not get modified by the addition of the weak gauge $U(1)_{R}$
interactions under which the two squark fields have charge $\tfrac{1}{2}$ and  $-\tfrac{1}{2}$, respectively.   $\nu$ is the $U(1)_{0}$ winding index depending on the gauge group considered: $U(1)_{0}\times G^{'}$,  $G'= SU(N), SO(2N), USp(2N)$, etc.~\cite{Eto:2008yi}. For $G'=SU(N)$,  $\nu= 
\tfrac{k}{N}$ where $k$ is the winding number of the vortex. 
As a result we expect, in our case, $N=N_{f}=2, \, k=1$,  the two zero modes  (the complex $\cp$ coordinate $B$) to persist somehow
upon weak $U(1)_{R}$ gauging.  

\subsection{The origin of the vortex moduli \label{calcul}} 

Following the familiar procedure~\cite{Eto:2006pg} we now set
\beq
Q(z,\bar z)=S_{L}^{-1}(z,\bar z)H_{0}(z)S_{R}^{-1}(z,\bar z)\;,  \qquad
A_{\bar z}= i S_{L}^{-1} \partial_{\bar z} S_{L}\;, \quad  A^{R}_{\bar z}= i S_{R}^{-1} \partial_{\bar z} S_{R}  \;, \label{mmm}
\eeq
\beq     S_{L} =  e^{\psi_{0}} \left(\begin{array}{cc}e^{\psi_3 /2} & 0  \\  e^{\psi_3/2}  \, w & e^{- \psi_3/2}\end{array}\right),  \quad 
S_{R} =  \left(\begin{array}{cc}e^{\psi_R  /2} & 0 \\  0    & e^{- \psi_R  /2}  \end{array}\right)\;,   \quad H_{0}=  \left(\begin{array}{cc}z & 0 \\ B  & 1\end{array}\right)\,, 
\label{Smatrices}   \eeq
where $B$ is a complex number and $z \equiv x+ i y$.\footnote{ $S_{L}$ can be and  has been chosen to have a lower triangular form by an appropriate $SU(2)_{L}$ gauge transformation. Also, the functions $\psi_{0}, \psi_{3}, \psi_R$ have been set to be real by an appropriate diagonal $U(1)_{0} \times SU(2)_{L} \times U(1)_{R}$ gauge rotation.   }
We also recall that the moduli matrix $H_{0}(z)$ and $S$ matrices in (\ref{mmm}) are defined up to a ``gauge choice'' of the form
\beq     H_{0}(z) \to   V_{L}  \, H_{0}(z) \, V_{R}(z)\,,\qquad  S_{L} \to   V_{L}  S_{L}\;,  \, \quad S_{R} \to  S_{R}  V_{R}\,,   \label{Vgauge}
\eeq
where the holomorphic matrices $V_{L}(z)$ and $V_{R}(z)$ belong to the complexifications of the $SU(2)_{L} \times U(1)$ and $U(1)_{R}$ gauge groups respectively. 
In the analysis of the vortex moduli spaces as complex manifolds, these $V$-equivalence relations play a fundamental role. 

~~~~

\noindent{ \bf Remarks}  

We continue to use here  the same letter  $B$  for the complex parameter characterizing the solutions, as in the model of Subsection~\ref{standard} without the $U(1)_{R}$ gauging,   {\it even though} the physical meaning will be different. The normalizable $\cp$ zero modes $B$ were massless Nambu-Goldstone bosons 
traveling along the vortex length in the absence of $U(1)_{R}$ gauge fields;  
on the other hand, 
with the $U(1)_{R}$ weak gauging, only Arg$(B)$ remains the true Nambu-Goldstone mode, but it is coupled with and eaten by the exact, asymptotically massless electromagnetic field which propagates in the $4D$ bulk, 
while $|B|$ will determine the (electro-)magnetic AB flux carried by the vortex. The change in the nature of the zero modes reflects  the fact that now the non-Abelian vortex is coupled to an exact $4D$ gauge mode whose direction is nontrivially oriented with respect to the underlying ($g_{R}=0$ theory) non-Abelian vortex orientation.

~~~~

Eqs.~(\ref{mmm}) solve the BPS equation (\ref{bps3}), while the other two BPS equations (\ref{bps1}), (\ref{bps2}) turn into the master equations
\begin{eqnarray}
&&  4\, \partial_{\bar z}\left(\Omega\partial_{ z}\Omega^{-1}\right)+g_{L}^{2}\left(    H_{0}\,\Omega^{-1}_{R}H_{0}^{\dagger}\Omega^{-1} -\xi   \right)=0\;,\qquad \Omega \equiv S \, S^{\dagger }\;,   \label{mastereq} \\
&&  4\, \partial_{\bar z}\left(\Omega_{R}^{-1}   \partial_{ z}\Omega_{R}   \right)+g_{R}^{2}\left( \Omega_{R}^{-1} H_{0}^{\dagger}\Omega^{-1} H_{0}  -   \frac{{\mathbbm 1}  }{2} 
  \Tr   ( \Omega_{R}^{-1}  H_{0}^{\dagger}\Omega^{-1} H_{0} )  \right)=0\;,\qquad \Omega_{R} \equiv S_{R}^{\dagger } \, S_{R}\;. \nonumber    \end{eqnarray}
According to Eq. (\ref{mmm}) the scalar field takes the form, 
\begin{eqnarray}
 Q&=&S_{L}^{-1} H_{0}(z) S_{R}^{-1} =\sqrt{\xi}e^{-\psi_0}\left(\begin{array}{cc}
     z e^{-\psi_3/2} & 0 \\ -(z w+B) e^{\psi_3/2}&e^{\psi_3/2}
	  \end{array}\right)\, e^{-\psi_R \,{\tau^{3}}/{2}}    \nonumber \\
&=&   \sqrt{\xi}e^{-\psi_0}\left(\begin{array}{cc}
     z e^{-\psi_+ /2} & 0 \\ -(z \tilde w +\tilde B ) e^{\psi_+  /2}&e^{\psi_+ /2}
	  \end{array}\right)\, ,   \label{squark}
\end{eqnarray}
where \footnote{For the sake of clarity we recall that  the functions $\psi_{+}$, $\psi_{3}$, and $\psi_{R}$    correspond to 
the gauge fields $B_{\mu}$, $A^{L\, 3}_{\mu}$ and  $A^{R}_{\mu}$ respectively.    }
\begin{eqnarray}
 \psi_+\equiv \psi_3+\psi_R\;,\quad 
\tilde w \equiv we^{-\psi_R}\;,\quad
\tilde B\equiv B e^{-\psi_R}\; .
\end{eqnarray}

 After the $SU(2)_{L}\times U_{0}(1)\times U(1)_{R}$ gauge freedom is used to fix the form the the $S$ matrices, Eq.~(\ref{Smatrices}), there remains still an arbitrariness of the $V_{R}$ transforms, (\ref{Vgauge}).  The latter can be used to fix the value of  $\psi_{R}(0)$, for instance, to $0$. This unambiguously defines the meaning of the parameter $B$. The physical meaning of the parameter $B$ can be understood as the ratio of the scalar fields of the first and second flavors at the vortex core ($z=0$).

Note that the electromagnetic gauge transformation acts on  $Q$  as
\beq   Q \to      \left(\begin{array}{cc}e^{-i \beta/2} & 0 \\0 & e^{i \beta/2}\end{array}\right)  \, Q\, \left(\begin{array}{cc}e^{i \beta/2} & 0 \\0 & e^{-i \beta/2}\end{array}\right)\;. \label{action}
\eeq
Keeping the functions $\psi_{0}, \psi_{3}, \psi_R$ real via a combination of the $V$ transformations,  $B$ can be seen to transform under the electromagnetic gauge transformations simply  as
\beq   B \to     e^{ i \beta}\, B\,.    \label{Bemtrans}
\eeq 

For a comparison with the physical approach of the previous section it is convenient to recall the relation between the gauge fields and  various $\psi$ functions
\beq    \bar A_{R}  \equiv   A^{R}_{x}- i A^{R}_{y}   =  - \frac{i}{g_{R}}   \bar \partial \psi_R  \;, \qquad   \bar A_{0}  \equiv     A^{(0)}_{x}- i A^{(0)}_{y}   = - \frac{i}{g_{0}}    \bar \partial \psi_0 \;;   \label{relations} \eeq
the $SU(2)_{L}$ gauge fields are related to $\psi$'s and $w$ through 
\beq  g_{L} \, A_{\bar z}= i S_{L}^{-1} \partial_{\bar z} S_{L}\;.  \eeq
Explicitly  Eq.~(\ref{mastereq}) becomes   
\beq    \frac{4}{g_{0}^{2}}  \, \de \bar \de \psi_{0} &=&  2 \xi -  \xi  e^{- 2 \psi_{0}} \left(   |z|^{2}  e^{-\psi_{3}- \psi_R  } + ( e^{\psi_R }  + | z w  + B  |^{2} e^{-\psi_R  }) \, e^{\psi_{3}}      \right) \;;  \nonumber \\
   \frac{4}{g_{L}^{2}}  \, \de \bar \de \psi_{3}  -   \frac{4}{g_{L}^{2}}  \, |\de \bar w |^{2} e^{2 \psi_{3}}   &=&   \xi  e^{- 2 \psi_{0}} \left(  -  |z|^{2}  e^{-\psi_{3}- \psi_R  } + ( e^{\psi_R }   + | z w  + B  |^{2}  e^{-\psi_R  } ) \, e^{\psi_{3}}    \right) \;;   \nonumber \\
\frac{1}{g_{L}^{2}}  \,  \de  ( 2 \, e^{2 \psi_{3}}  {\bar \de}  w ) &=&    \xi \, {\bar z}  ( z w  + B  )\, e^{- 2\psi_{0}+ \psi_{3}- \psi_R } \;;  \nonumber \\
  \frac{4}{g_{R}^{2}}  \, \de \bar \de \psi_R  &=&   \xi  e^{- 2  \psi_{0}} \left(  -  |z|^{2}  e^{-\psi_{3}- \psi_R  } + ( e^{\psi_R }   -  | z w  + B  |^{2}  e^{-\psi_R  }) \, e^{\psi_{3}} \right)\;. \label{mastereqBis}
\eeq 
Let us first consider the $B=0$ solution.  By setting
\beq  B =0\;, \qquad  w=0\;, \qquad    \psi_R  /  {g_{R}^{2}} =    {\psi_{3}} /  {g_{L}^{2}}\;,  
\eeq
the master equations reduce to the first two,  which are exactly the equations for the  familiar  non-Abelian vortex with the $B=0$ orientation. This solution was discussed in Subsection~\ref{BPSB0}. 

The solutions with $B \ne 0$ are more interesting.  Note that by redefining
\beq   w  \to  B \, {w}\,, \qquad   \psi_R  \to  \psi_R  +   \log  |B|\,, \qquad     \psi_3 \to   \psi_3  -   \log  |B| \;,
\eeq
the parameter $B$ can be eliminated  from all four equations (\ref{mastereqBis}) and gets replaced by $1$ everywhere. It might thus look as if the set of equations (\ref{mastereqBis}) involved a redundant set of functions and that there is actually only one solution for $B \ne 0$, equivalent to the $B=1$   solution discussed in  Subsection~\ref{BPSB1}.

Actually,  this is not so. 
The coupled differential equations (\ref{mastereqBis})  define the solution only after a specific set of boundary conditions are imposed. 
The appropriate boundary conditions at large $r$  are  
\begin{eqnarray}
\psi_0\sim \frac 12 \log|z|\;,\quad \psi_3+ \psi_{R} \sim \log|z|\;,\quad
  w\sim -\frac{B}{z}\;,       \label{boundcond}
\end{eqnarray}
so that the squark fields approach the form, 
\beq   Q   \sim    \sqrt{\xi} \,  \left(\begin{array}{cc}e^{i \varphi}  & 0 \\0 & 1\end{array}\right)\;.    \label{KOgauge}
\eeq
But this does not uniquely fix the boundary conditions for $\psi_{ 3}$ and $\psi_{R}$ separately: there remains a freedom in sharing $\log |z|$ between them.

A general boundary condition is thus
\begin{eqnarray}
  \psi_{R} \sim  \delta   \log|z|+\ldots\;,\quad \psi_3\sim
  (1-\delta)\log|z| +\ldots \;.\end{eqnarray}
Furthermore if  we define \footnote{ Recalling that $\psi$ functions contain the coupling constants, it can be seen that this combination  corresponds exactly to $A^{em}$  of Eq.~(\ref{em}).} 
\begin{eqnarray}
 \psi^{em} \equiv\frac1{g_{L}^2+g_{R}^2}\left(g_{L}^2 \psi_R- g_{R}^2\psi_3\right)\;,
\label{gamma} 
\end{eqnarray}
one has 
\begin{eqnarray}
 \psi_+ = \psi_{3}+ \psi_{R}  \sim \log|z|\;,\qquad  \psi^{em}  \sim    (\delta -   \frac{g_{R}^{2}}{g_{L}^{2}+ g_{R}^{2}}) \,  \log|z|\;.  \label{boundarycond}
\end{eqnarray} 
In the next section we solve Eqs.~(\ref{mastereqBis})  explicitly for small $g_{R}$ and determine  $\delta$:  the asymptotic behaviors of $\psi_{R}$,  $\psi_{+}$, $\psi_{3}$
and $\psi_{0}$\footnote{Notice here that the invariance of the master equation (\ref{mastereqBis}) under rescaling of the modulus of $B$ can be regarded as a consequence of the fact that our vortex solutions are BPS saturated, thus preserving half of the supersymmetries when our model is embedded in an $\mathcal N=2$ theory. In supersymmetric gauge theories, gauge symmetry is naturally extended to a complex symmetry. The phase symmetry in Eq.~(\ref{Bemtrans}) of the parameter $B$ thus implies a symmetry under rescaling of the modulus of $B$. As discussed, however, boundary conditions are sensitive to the modulus of $B$, and this is responsible for the existence of a nontrivial set of (gauge inequivalent) vortex solutions.}.

\subsection{Perturbative expansion in  $\lambda =  g_{R}/g_{L}$   \label{perturb} }

We will now solve Eq.~(\ref{mastereqBis}) perturbatively in $\lambda \equiv  g_{R}/g_{L}$.   At zeroth order one must solve the equations for generic $B$ 
at $g_{R}=0$.   Setting $g_{R}=0$ one has 
\begin{eqnarray}
 Q&=&S_{L}^{-1} H_{0}(z)   =\sqrt{\xi}e^{-\psi_0}\left(\begin{array}{cc}
     z e^{-\psi_3/2} & 0 \\ -(z w+B) e^{\psi_3/2}&e^{\psi_3/2}
	  \end{array}\right)\;,
 \label{squark0}
\end{eqnarray}
and the functions appearing here satisfy
\beq    \frac{4}{g_{0}^{2}}  \, \de \bar \de \psi_{0} &=&  2 \xi -  \xi  e^{- 2 \psi_{0}} \left(   |z|^{2}  e^{-\psi_{3}} + ( 1  + | z w  + B  |^{2}) \, e^{\psi_{3}}      \right) \;;  \nonumber \\
   \frac{4}{g_{L}^{2}}  \, \de \bar \de \psi_{3}  -   \frac{4}{g_{L}^{2}}  \, |\de \bar w |^{2} e^{2 \psi_{3}}   &=&   \xi  e^{- 2 \psi_{0}} \left( - |z|^{2}  e^{-\psi_{3} } + (1  + | z w  + B  |^{2}   ) \, e^{\psi_{3}}    \right) \;;   \nonumber \\
\frac{1}{g_{L}^{2}}  \,  \de  ( 2 \, e^{2 \psi_{3}}  {\bar \de}  w ) &=&    \xi \, {\bar z}  ( z w  + B  )\, e^{- 2\psi_{0}+ \psi_{3}} \;.
 \label{mastereqBisbis}
 \eeq
We first review the  $B=0$ solution,  $\psi_0^{(0)}\equiv \psi_0|_{B=0},\psi_3^{(0)}\equiv \psi_3|_{B=0}$,  $w=0$. In this case  
\beq    \frac{4}{g_{0}^{2}}  \, \de \bar \de \psi_{0}^{(0)} &=&  2 \xi -  \xi  e^{- 2 \psi_{0}^{(0)}} \left(   |z|^{2}  e^{-\psi_{3}^{(0)} } +   e^{\psi_{3}^{(0)}}      \right) \;;  \nonumber \\
   \frac{4}{g_{L}^{2}}  \, \de \bar \de \psi_{3}^{(0)}   &=&   \xi  e^{- 2 \psi_{0}^{(0)}} \left( - |z|^{2}  e^{-\psi_{3}^{(0)} } +  e^{\psi_{3}^{(0)}}    \right) \;. 
 \label{mastereqBisbisbis}
\eeq
To simplify the formulas somewhat,   below we will set
\beq      g_{0}^{2} = \frac{g_{L}^{2}}{2}\;;   \qquad    2  g_{L}^{2} \xi=1  \;,   
\eeq
the latter being simply the choice of the mass unit.  We then go to the singular gauge via the change of the variables
\beq         \psi_{0}^{(0)}   =   {\hat \psi}_{0}^{(0)}  + \frac{1}{2}  \log |z|\;, \qquad  \psi_{3}^{(0)} =      {\hat \psi}_{3}^{(0)}  + \log |z|\;, \label{singular}
\eeq
so that the equations become 
\beq    8  \, \de \bar \de {\hat \psi}_{0}^{(0)} &=& \frac{1}{2} \left[  2-   e^{- 2 {\hat \psi}_{0}^{(0)}} \left( e^{-{\hat \psi}_{3}^{(0)} } +   e^{{\hat \psi}_{3}^{(0)}}      \right) \right]+ 2 \pi \delta^{2}(x) \;;  \nonumber \\
   4  \, \de \bar \de {\hat \psi}_{3}^{(0)}   &=&    \frac{1}{2} \left[  e^{- 2 {\hat \psi}_{0}^{(0)}} \left( - e^{-{\hat \psi}_{3}^{(0)} } +  e^{{\hat \psi}_{3}^{(0)}}    \right)\right]  + 2 \pi \delta^{2}(x) \;.  
 \label{Taubes}
\eeq
The difference between the two equations yields 
\beq      2 \,  {\hat \psi}_{0}^{(0)}(z)   - {\hat \psi}_{3}^{(0)}(z) \equiv 0\;,   \label{B=0sol}
\eeq
whereas the sum yields Taubes'  equation \cite{Taubes:1980}
\beq    4\,    \de \bar \de   \varphi(z)  =  \frac1r  \frac{d}{dr}  (r   \frac{d}{dr}\varphi (z))     =  \frac{1}{2} (1 -   e^{- 2\varphi(z)} ) +  2 \pi \, \delta^{2}(x)\;,   \label{Taube}
\eeq
for 
\beq    ( 2\, {\hat \psi}_{0}^{(0)}(z) +   {\hat \psi}_{3}^{(0)}(z))/2   =  2 {\hat \psi}_{0}^{(0)}(z) =  {\hat \psi}_{3}^{(0)}(z) \equiv \varphi(z) \;,  \label{varphi}   \eeq
with the boundary conditions
\beq   \lim_{r\to 0} r \varphi'=-1\;, \qquad \lim_{r\to \infty}\varphi=0\;.  \label{TaubeBC}
\eeq
 The solution  $\varphi(r)$ behaves as
\begin{eqnarray}
 \varphi(r)  =\left\{
\begin{array}{cc}
 -\log r +a & {\rm for~} r\ll 1 \\
q\, K_0(r)  & {\rm for~} r \gg 1
\end{array}\right\} \,, \label{Taubesol}
\end{eqnarray}
with constants~\cite{MinoruKeisuke}  
 \beq a=0.50536..., \qquad    q=1.707864....\label{achoice}    \eeq  
  The modified Bessel function of the second kind $K_0$ is exponentially damped at large $r$. 
This gives $Q|_{B=0}.$

The solutions at $B \ne 0$ can be found by a color-flavor rotation of $Q|_{B=0}$,
\beq     Q =   U  \, Q|_{B=0}   \, U^{-1},  \qquad   U=  \frac{1}{\sqrt {|B|^{2}+1}}    \left(\begin{array}{cc}1 & -B^* \\B & 1\end{array}\right)\;, 
\eeq
followed by an appropriate  $SU(2)_{L}$ gauge transformation to bring $Q$ back to the form,  (\ref{squark0}).   The answer is (see Appendix~\ref{solutions}): 
\begin{eqnarray}
\psi_0&=&\psi_0^{(0)}  \;;  \nn
 \psi_3&=&\psi_3^{(0)}
+\log\left\{\frac{1+|z|^2|B|^2e^{-2 \psi_3^{(0)}}}{1+|B|^2}\right\}\;;    \nn  
w&=&-\bar z B e^{-\psi^{(0)}_3-\psi_3} =   \frac{-\bar z B (1+|B|^2)e^{-\psi_3^{(0)}}}
{e^{\psi^{(0)}_3}+|B|^2|z|^2e^{-\psi^{(0)}_3}}\;.  \label{SolB}
\end{eqnarray}

Once one knows these solutions  with $\lambda=  g_{R}/g_{L}= 0$ for generic $B$,  the solutions for 
  $\lambda \ll 1$ can be found by perturbation theory.  Using Eq.~(\ref{SolB}), the fourth BPS equation (\ref{mastereqBis}) yields 
\beq   \frac{4}{g_{R}^{2}}  \, \de \bar \de \psi_R  &\simeq&   \xi \,   e^{- 2  \psi_{0}} \left(  -  |z|^{2}  e^{-\psi_{3}} + ( 1   -  | z w  + B  |^{2}) \, e^{\psi_{3}} \right)\;
\nonumber \\
&=&  -  \xi \, \frac{|B|^{2}-1 }{|B|^{2}+1 }  (1 -  e^{-2 \varphi })=   -  \frac{1}{2 g_{L}^{2}}\, \frac{|B|^{2}-1 }{|B|^{2}+1 }  (1 -  e^{-2 \varphi })  \;.    
\eeq
Next, using the Taubes' equation (\ref{Taubes}) for $\varphi$, one finds
\beq
 \frac1r (r \psi_R' )'   =-\frac{g_{R}^{2}}{g_{L}^{2}}\,  \frac{|B|^2-1}{|B|^2+1}
\times \frac1r (r \varphi')'  +{\cal O}(\lambda^2) \;;
\eeq
this means that  
\begin{eqnarray}
 \psi_R(r) =-\frac{g_{R}^{2}}{g_{L}^{2}}\,  \frac{|B|^2-1}{|B|^2+1} 
\left(\varphi(r) +\log  r  - a  \right)+{\cal O}(\lambda^2)    \label{final}
\end{eqnarray}
where we have recalled Eq.~(\ref{TaubeBC}) and have appropriately taken into account the boundary conditions
for $\psi_R$\footnote{Note that  $\varphi =  {\hat \psi}_{3}^{(0)}(z)$ was defined in the singular gauge, see Eqs.~(\ref{Taubesol}) and~(\ref{achoice}),   whereas   $\psi_R$ represents the original right gauge field (Eq.~(\ref{relations})) and therefore is regular at the vortex core.   $\psi_{R}(0)$  can be set to  $0$ without losing generality,  by an appropriate choice of the  $V$-gauge (Eq.~(\ref{Vgauge})). }.

The large $r$ behavior 
\begin{eqnarray} 
  \psi_{R}\sim  \delta \log r, \qquad   \delta = -\frac{g_{R}^{2}}{g_{L}^{2}} \frac{|B|^2-1}{|B|^2+1} \;\label{asymp2}  \end{eqnarray}
follows thus from Eq.~(\ref{Taubesol}).   Recalling  (Eq.~(\ref{boundcond}), Eq.~(\ref{KOgauge}))   that $\psi_{+}= \psi_{3}+ \psi_{R}\sim \log r$,  one finds to this order  that 
\beq  \psi_{3} \sim  (1-\delta) \log r =   (   1 +   \frac{g_{R}^{2}}{g_{L}^{2}} \frac{|B|^2-1}{|B|^2+1} )   \log r\; .
\eeq
For completeness,   we get for $\psi^{em}$  (Eq.~(\ref{boundarycond})) 
\beq  \psi^{em}   \sim    (\delta -   \frac{g_{R}^{2}}{g_{L}^{2}+ g_{R}^{2}}) \, \log r    \sim - \frac{g_R^2}{g_L^2}  \frac{2 |B|^{2}}{|B|^{2}+1} \, \log r\;, \eeq
whereas the $U_0(1)$ gauge fields winds half  (E.~({\ref{boundcond})): 
\beq     \psi_{0} \sim \frac{1}{2} \log r\;. 
\eeq
The asymptotic behavior of  the gauge fields  $A^{L\,3}_i$ and  $A_i^{R}$ is then
\begin{eqnarray}
 A^{L\,3}_i= - \frac1{g_{L}} \epsilon_{ij}\partial_j   \psi_{3} \sim   - \frac{1-\delta} {g_{L}} \epsilon_{ij}  \frac{x_j }{r^2}\;,\qquad
 A_i^{R} = - \frac1{g_{R}} \epsilon_{ij}\partial_j   \psi_{R}  \sim  -  \frac{\delta}{g_{R}} \, \epsilon_{ij}  \frac{x_j }{r^2}    \;.     \label{asymp1}
\end{eqnarray}
For completeness, the ``broken'' gauge field $B_{\mu}$ behaves asymptotically as
\beq     B_{i}= - \frac1{g^{\prime}}  \epsilon_{ij}\partial_j   \psi_{+} \sim   \frac1{g^{\prime}}  \epsilon_{ij} \frac{x_j }{r^2}\;.   \label{brokenBis}
\eeq
This leads, for a particle carrying only the unit charge with respect to $U(1)_{R} $, to an AB phase
 \beq   g_{R}\oint  dx_{i}   A^{R}_{i} =  
 g_{R}\int  d^{2}x \, F^{R}_{12} =  
   \frac{2 \pi g_{R}^{2}}{g_{L}^{2}}   \frac{ |B|^{2}-1}{ |B|^{2}+1}\;,   \label{simpleBis}
 \eeq
 as it goes around the vortex, 
in accordance with  Eq.~(\ref{simple});  
one sees  that the three explicit solutions we found earlier, with
$|B|=0,1,\infty$,  are interpolated by the modulus $|B|$.

\section{Aharonov-Bohm effect} 

    The result (\ref{simpleBis}) is well defined and gauge invariant. A particle  with a unit $U(1)_{R}$ charge but with no charges 
 with respect to the $SU(2)_{L}\times U_{0}(1)$ will experience the AB effect (\ref{simpleBis}). More generally, a particle in a definite representation of the
 $SU(2)_{L}\times U_{0}(1) \times U(1)_{R}$  gauge group will get a definite AB phase after encircling around the vortex.  Some examples are shown in Table \ref{tabella}.
 We note that, in the case of the  $U(1)_{R}$ charge one particle (particle K in Table \ref{tabella}), the AB phase is maximum (in the magnitude) for the vortices $B=0$ and $B= \infty$, where the vortex orientation and the external gauging are aligned, whereas for the vortices $|B|=1$ (the vortex solutions along the equator of $CP^{1}\sim S^{2}$) the orientations are orthogonal and there is no AB effect.

    As the $U(1)_{R}$ gauge symmetry is a spontaneously  broken symmetry,  it might be thought that  
any physical effect at large distances (and far from the vortex core) should be describable in terms of the coupling of the particle to the massless gauge  field, $A^{em}_{\mu}$.   For instance, a particle carrying a unit charge with respect to  $U(1)_{R}$ (but with no other charges)  interacts through the covariant derivative
\beq    (\partial_{\mu}  +  \frac{i}{2}    g_{R}  A^{R}_{\mu}  ) \, K =   \left(\partial_{\mu}  +  \frac{i}{2} e  A^{em}_{\mu}   +  \frac{i}{2}   \frac{g_{R}^{2}}{\sqrt{g_{L}^{2}+ g_{R}^{2}}}  B_{\mu} \right) \, K \;.  
\eeq
It would seem natural to assume that the long-distance physics is dominated by the coupling to the massless ``photon'' field $A^{em}_{\mu}$,  the interactions with the massive $B_{\mu}$ field providing some small corrections  (the ``weak interactions'')  calculable in perturbation theory. 

In the presence of vortex, this is not quite  so. As one can explicitly see from (\ref{brokenBis}) also the massive gauge fields contribute to the AB effect to the same order, as is well known \cite{Alford:1988sj}. As a result, two particles with the same electromagnetic charge but with different couplings  to the broken gauge field $B_{\mu}$ such as $K$ and  $\psi_{1}$ in Table~\ref{tabella} (i.e., two particles belonging to different representations of  $SU(2)_{L}\times U_{0}(1) \times U(1)_{R}$) experience different AB effect going around the vortex, however far from it.

    One must also be somewhat careful in deriving the physical AB effect  from the results of the calculations in the preceding sections,  as one is working with a spontaneously broken $SU(2)_{L}$ gauge theory. The situation is somewhat analogous to that of the Weingberg-Salam theory.  In the Weinberg-Salam theory  the neutrino and electron are usually described by the upper and lower components of an $SU_{L}(2)$ doublet field; nevertheless both describe physical particles of definite mass and charge.  Of course, the solution of this apparent puzzle is well known:  the identification of 
the two leptons  with the two components of a doublet is correct  in a gauge in which the upper component of the Higgs doublet gets the  nonvanishing VEV. 
More approariately, the electron and neutrino must be associated with some appropriate $SU_{L}(2)$ gauge-invariant composite fields involving the left-handed lepton and the Higgs scalar fields\cite{THCargese}. 

In our case,    $SU(2)_{L}\times U_{0}(1) \times U(1)_{R}$ gauge symmetry is broken  in the bulk  by the scalar VEVs,
\beq
\langle Q \rangle =  \sqrt {\xi}
\left(
\begin{array}{cc}
1  & 0    \\
  0 & 1
\end{array}
\right)   
\label{vacuumagain}  \eeq
 (see  Eq.~(\ref{vacuumbis}))   to $U^{em}(1)$.  The identification of 
\beq  A_{\mu}^{em}=   \frac{g_L}{\sqrt{g_{L}^{2}+ g_{R}^{2}}} A_{\mu}^{R} - \frac{g_R}{\sqrt{g_{L}^{2}+ g_{R}^{2}}}   A_{\mu}^{L\, 3} \label{emagain} \eeq
with the massless ``photon'' (and $B_{\mu}$ with the massive field, see Eqs.~(\ref{em})-(\ref{inverse})) is correct in the gauge in which the scalar VEVs take the form, 
(\ref{vacuumagain}).

The vortex  solutions also depend on the gauge used to solve the field equations.  The solution in Subsections \ref{calcul}, \ref{perturb}  has been obtained by working in the gauge where the scalar fields have the asymptotic form 
\beq   Q   \sim    \sqrt{\xi} \,  \left(\begin{array}{cc}e^{i \varphi}  & 0 \\0 & 1\end{array}\right)\;   \label{KOgaugebis}
\eeq
 (Eq.~(\ref{KOgauge})).   $A_{i}^{R}$ and $A_{i}^{L\, 3}$ are found to behave as in (\ref{asymp1}), (\ref{asymp2}) in this gauge.  Note that the identification of  the combination  (\ref{emagain}) with the masseless field in the bulk is appropriate in this gauge also \footnote{We thank Chandrasekhar Chatterjee for discussions on this point.}.    
 
Even if choosing a gauge is unavoidable as in any other gauge theory calculations,  clearly the concept of the AB phase which
a particle in any definite representation  of the $SU(2)_{L}\times U_{0}(1)\times U(1)_{R}$ group will acquire in encircling the vortex far from it,  is  a physical one.  
The only unusual aspect is that, as pointed out above,  the AB effect does  not 
depend only on the ``electromagnetic'' charge but also on the ``weak'' charge.    The calculation can be done in any gauge, of course, but when the result is transformed back to the conventional gauge where the scalar VEV's approach a  color-flavor unit matrix form, in order to relate it to the physical effect, the answer is 
the same. 

As a further remark on the gauge invariant nature of our results,  let us note that, as 't Hooft's observation for the electroweak theory \cite{THCargese},  a particle described by a gauge variant field such as $\psi_{1}$ (in our conventional gauge) can  be regarded as a gauge invariant object associated with an $SU(2)_{L}$ singlet composite field such as   
${\bar  Q^{(1)}}  \psi $ or  $\epsilon_{ij} Q^{(2)}_{i} \psi_{j}$.  The physical AB phase is indeed the same in all these descriptions, see Table~\ref{tabella}.

\begin{table}[h]
  \centering 
  \begin{tabular}{|c|c|c|c|c|c|c|}
\hline
 Particle  &  $ U(1)_{L3}   \subset SU(2)_{L}  $ & $U(1)_{0}$  &   $U(1)_{R}$  & $U^{em}(1)$ & $U_{B}(1)$ &   AB phase  \\
   \hline
   $K$     & $~~ 0  \qquad   {\underline  1}$   & 0  &    $1$   &  $ \tfrac{1}{2}  $ &  $\tfrac{\gamma}{2}$ & $- \tfrac{\delta}{2}$  \\
    $L$     & $~~ 0  \qquad  {\underline 1}$   & 0  &   $ -1 $  &  $ - \tfrac{1}{2}$ &    $-\tfrac{\gamma}{2}$  &   $ \tfrac{\delta}{2}$ \\
     $\psi_1$     & $ ~~ 1 \qquad    {\underline 2}$   & $0$  &   $0$  &  $ -\tfrac{1}{2} $&  $\tfrac{1-\gamma}{2}$   &   $\tfrac{\delta-1}{2}$    \\
          $\psi_2$     &  $ -1 \qquad {\underline 2}$   & $0$  &   $0$  &  $\tfrac{1}{2} $     &  $- \tfrac{1-\gamma}{2}$   & $\tfrac{1-\delta}{2}$   \\
               $\chi_1$     & $ ~~ 1 \qquad    {\underline 2}$   & $0$  &   $1$  &  $ 0$&  $\tfrac{1}{2}$   &   $-\tfrac{1}{2}$    \\
          $\chi_2$     &  $ -1 \qquad {\underline 2}$   & $0$  &   $1$  &  $1 $     &  $- \tfrac{1}{2}$   & $\tfrac{1}{2}-\delta$   \\
          \hline
  $Q^{(1)}_{1}$   & $  ~~ 1 \qquad  {\underline 2}$  &  $1$  &  $ 1$ &   $0$  &  $ \tfrac{1}{2}  $    &   $0$   \\
 $Q^{(1)}_{2}$     & $ -1 \qquad{\underline 2}$   & $1$  &   $ 1 $  &  $1$   &     $\gamma  $   &   $ - \delta$   \\
  $Q^{(2)}_{1}$   & $  ~~ 1 \qquad  {\underline 2}$  &  $1$  &  $  -1$ &  $-1$ &  $ - \gamma$      &   $ + \delta$  \\
 $Q^{(2)}_{2}$     & $ -1 \qquad{\underline 2}$   & $1$  &   $ -1 $  &  $0$    &  $-\tfrac{1}{2}$    &   $0$   \\
 \hline
  ${\bar  Q^{(1)}}  \psi $   &   $ ~~ 0  \qquad  {\underline 1}$  &  $-1$  &  $  -1$ & $ - \tfrac{1}{2}$  &  $-\tfrac{\gamma}{2}$  &   $\tfrac{\delta -1}{2}$   \\
 $\epsilon_{ij} Q^{(1)}_{i} \psi_{j}$     & $~~ 0  \qquad {\underline 1}$    & $1$  &   $ 1 $  & $ \tfrac{1}{2}$    &  $\tfrac{\gamma}{2}$     &   $\tfrac{1 -\delta }{2}$    \\
 ${\bar  Q^{(2)}}  \psi $   &   $  ~~ 0  \qquad {\underline 1}$  &  $-1$  &  $ 1$ & $ \tfrac{1}{2}$  & $\tfrac{\gamma}{2}$  &   $-\tfrac{\delta +  1 }{2}$  \\
 $\epsilon_{ij} Q^{(2)}_{i} \psi_{j}$     & $~~ 0  \qquad  {\underline 1}$    & $1$  &   $ -1 $  &   $  -\tfrac{1}{2}$    &  $-\tfrac{\gamma}{2}$   &    $\tfrac{\delta +  1}{2}$   \\
\hline
  ${\bar  Q^{(1)}}  \chi$   &   $ ~~ 0  \qquad  {\underline 1}$  &  $-1$  &  $  0 $ & $ 0 $  &  $0$  &   $-\tfrac{1}{2}$   \\
 $\epsilon_{ij} Q^{(1)}_{i} \chi_{j}$     & $~~ 0  \qquad {\underline 1}$    & $1$  &   $ 2 $  & $ 1 $    &  ${\gamma}$     &   $\tfrac{1 }{2}- \delta$    \\
 ${\bar  Q^{(2)}}  \chi $   &   $  ~~ 0  \qquad {\underline 1}$  &  $-1$  &  $ 2$ & $ 1$  & ${\gamma}$  &  $ -\tfrac {1 }{2}-\delta $    \\
 $\epsilon_{ij} Q^{(2)}_{i} \chi_{j}$     & $~~ 0  \qquad  {\underline 1}$    & $1$  &   $ 0 $  &   $ 0 $    &  $0$   &   $\tfrac{1 }{2}$      \\
\hline
\end{tabular}
  \caption{$ \delta = -\frac{g_{R}^{2}}{g_{L}^{2}} \frac{|B|^2-1}{|B|^2+1};$   $\gamma \equiv \tfrac{g_{R}^{2}}{g_{L}^{2}+ g_{R}^{2}}$.     In the table are some fields with their charges under the various gauge groups, and the associated AB phase  (given in the unit of $2 \pi$ and {
  \it modulo} $2\pi$, so that $- \tfrac{1}{2} \sim \frac{1}{2}$).      The extra factor $\tfrac{1}{2}$'s  in the last three columns take into account of the $\tau^{3}/2$ in front of the $A^{L,3}_{\mu}$ and  $A^{R}_{\mu}$.    The electromagnetic charge represents the coupling to  $A^{em}_{\mu}$ in the unit of the coupling constant $e$;  the coupling to the broken
  gauge field $B^{\mu}$ is  in the unit of $g^{\prime}=\sqrt{g_{L}^{2}+ g_{R}^{2}}$.  
    }\label{tabella}
\end{table}

\section{Low-energy effective action}

When the moduli parameters $B$ are allowed to fluctuate along the vortex length and in time, i.e., in the vortex worldsheet, 
the associated collective coordinates become dynamical, described by a long-wavelength effective action. In the model of Section 1 (without the $U(1)_{R}$ gauging) this is just a $2D$ $\cp$ sigma model, Eq.~(\ref{CP1sigma}), with K\"ahler potential~\cite{Hanany:2003hp} --\cite{GJK},
\beq   K^{(CP)}=   ({4\pi}/{g_{L}^2})   \log  (|B|^{2}+1)\;, \label{CP1Kahler} \eeq
corresponding to the Fubini-Study metric. 

In the presence of $4D$ bulk zero modes -- an exact unbroken $U(1)^{em}$ gauge symmetry --  coupled to the vortex dynamics, the straightforward approach of Appendix A cannot be applied. Although there are nontrivial mixings of $A^{em}$ with other broken gauge fields inside of the vortex, the vortex configurations themselves can be approximated by setting  $g_{R}=e=0$  to  lowest oder.  One then finds~\cite{KNV} 
\beq    {\cal L}_{eff}  =  - \frac{1}{4}  \int d^{4}x   \,   (F^{em}_{\mu \nu})^{2} 
+   \frac{4 \pi}{g_{L}^{2}}   \int dt dz    \frac{1}{( |B|^{2}+1)^{2} }  \nabla_{\alpha}  {\bar B}  \nabla_{\alpha}   B  +  O(e^{3})\;,   \qquad  \alpha =3,0 \label{EffAc}
\eeq
where 
\beq    \nabla_{\alpha}   B =   (\de_{\alpha}  + i e A_{\alpha}^{em})  B\;. 
\eeq
This form is dictated by electromagnetic gauge invariance, under which the $B$ and $A^{em}$ fields transform as (see Eq.~(\ref{Bemtrans})): 
\beq  B \to  e^{ i \beta}  B\;,  \qquad  A_{\alpha}^{em} \to A_{\alpha}^{em} - \frac{1}{e}   \de_{\alpha} \beta\;.
\eeq
The current carried along the vortex is given by 
\beq   J_\alpha =
  -\frac{8\pi e }{g_{L}^2}\frac{\bar B \nabla_\alpha B-B
\nabla_\alpha \bar B}{2i (1+|B|^2)^2}+{\cal O}(e^3) \label{current}
\eeq
and  is proportional to 
\beq      \frac{e^{2}}{g_{L}^{2}} \,  \frac{|B|^{2}}{(|B|^{2}+1)^{2}} \;, 
\eeq
which is conserved along the vortex, 
\beq  \de^{\alpha}  J_{\alpha}  =0\;.   
\eeq

Dynamical aspects of the vortex zero modes (their fluctuations)  are subtle, as the $2D$ vortex zero modes  interact nontrivially with the $4D$ bulk zero modes (the ``photon'').   Eq.~(\ref{EffAc}) appears to give rise to the familiar Higgs mechanism, giving the photon a small nonvanishing  mass,
\beq   m_{\gamma}^{2}   \sim \frac{e^{2}}{g_{L}^{4}} \, \frac{|B|^{2}}{ (|B|^{2}+1)^{2} }\;.
\eeq
However the $B$ condensate  lives only inside the vortex, whereas the photon is massless in the $4D$ bulk outside the vortex.  As the action is quadratic in $A_{\alpha}$  the effect of integrating out the photon field can be determined.  For instance, a constant time variation  $\partial_{0} B$ would give rise to the charge density along the vortex:
\beq      q \sim     \frac{8 \pi  e}{g_{L}^{2}} \, \frac{|B|^{2}}{ (|B|^{2}+1)^{2} } \,  \partial_{0} {\rm Arg}  B\;.
\eeq 
The electromagnetic potential of an infinite string of such charge density is given by
\beq   A_{0}   \sim   \frac{q}{2\pi}  \log r\;,
\eeq
so that
\beq    \nabla A_{0} =  {\hat r} \partial_{r} A_{0} \sim  {\hat r} \frac{  q }{  2\pi   r}\;. 
\eeq
Substituting this into Eq.~(\ref{EffAc}) one finds a  (divergent)  energy 
\beq \frac{1}{2}  \int d^{3}x \, ( \partial_{i} A_{0} )^{2}  \sim    \frac{1}{2}   \int dz  \,  2 \pi  \int \frac{dr}{r}  \frac{ q^{2}}{4\pi^{2}}  =    \log \Lambda  \,    \int dz  \,    \frac{16 \pi  e^{2}}{g_{L}^{4}} \, \frac{|B|^{2}}{ (|B|^{2}+1)^{4} } \, \partial_{0} {\bar B}  \partial_{0} B\;
\eeq 
where an infrared cutoff in the transverse  plane at $\sqrt{x^{2}+ y^{2}}= \Lambda$ has been introduced.   

This physical discussion explains the result which one would obtain if one were to formally apply the standard method of calculation to the effective action in the presence of a massless $4D$ field:  one would find that the K\"ahler potential is given by  (Appendix~(\ref{effeAction}))
\beq K =  K_{core} + K_{bulk}\;, \eeq \beq 
K_{core} \simeq  K^{(CP)}=   \frac{4\pi} {{g_{L}^2}}   \log  (|B|^{2}+1)\;;  \qquad  \frac{\partial^{2}  K_{bulk}}{\partial B \partial {\bar B}} =   
 \frac{16\pi   e^{2}}{g_{L}^{4}}     \frac { |B|^{2}}{(|B|^{2}+1)^{2}} \, \log \Lambda \;, 
\eeq
namely a finite part which coincides (in the limit $g_{R}=0$) with the $\cp$ K\"ahler potential and a divergent part. As the above discussion shows,  
the latter is caused by the coupling of $B$ with the $4D$ massless modes, and it is such an interaction that effectively makes the $|B|$ mode non-normalizable.  

As a potential physical application, one may consider a vortex loop of finite size, as in Witten's cosmological string model~\cite{Witten} 
or a vorton \cite{Davis:1988jq},\cite{Davis:1988ij},
 and study the resulting finite physical effects.   We shall leave the study of these issues to a separate work. 

\section{Conclusion}

The AB effect found above is a result of the mismatch between the 
 the fixed $\tau^{3}_{R}$  weak gauging  direction  and the generic vortex orientation $B$ in the color-flavor $SU(2)$ space. 
 A particle with unit $U(1)_{R}$ charge, such as those in Table \ref{tabella},  will get an AB phase   
 as it travels along a large circle around the vortex.
 
 Although we restricted ourselves in this paper  to the simplest non-trivial prototype model based on $SU(2)_{L}\times U_{0}(1) \times U(1)_{R}$ gauge symmetry for the sake of  clarity of presentation,  it is indeed quite straightforward and rather interesting, to extend our analysis to more general gauge groups and patterns of partial weak gauging  (in preparation).   Also, even though our derivation and  the persistence of the  vortex moduli space upon $U(1)_{R}$ gauging both depend on the BPS nature of the model considered, the occurrence of the ``electromagnetic''  AB effect itself  has a clear physical explanation, and is independent of the BPS approximation.

 As noted in the Introduction, weakly gauging a $U(1)$ subgroup 
of the flavor symmetry occurs  in the color-locked phase of dense quark matter. 
Therefore, non-Abelian vortices in such a phase should also 
possess AB fluxes.   If such a CFL phase is realized in the cores of neutron stars,  
non-Abelian vortices will be created by a rapid rotation. 
Consequently, there may be  significant  AB effects on  particles charged under the asymptotically unbroken gauge symmetry, 
which could give considerable effects on evolutions of neutron stars. 
 In a non-BPS set-up, the tension of the vortex will generically depend on the value of the modulus $|B|$, and the magnitude of the AB effect will depend upon the which value of $|B|$ corresponds to the vortex solution with lowest tension. In Ref.~\cite{Vinci:2012mc} it was shown, in the case of high density QCD, that the vortex solutions with the smallest tension are those corresponding to $|B|=0,\infty$, which, as shown in this paper, have a non-trivial AB effect for the unbroken electromagnetic field.

In conclusion, as a result of the AB effect an external weak gauging $G_{W}$ of part of the exact color-flavor symmetry converts some of the 
internal orientational moduli of a non-Abelian vortex  into a new observable:  
an AB phase associated with $G_{W}$.  
This can be regarded as a novel physical property of non-Abelian vortices. 
The value of the phase depends on the particular solution considered, even though in our BPS systems the vortex tension is independent of the solution.

The extension of our results to more general gauge groups and weak gauging subgroups, including supersymmetric systems and roles of the fermions,  will be discussed in a forthcoming paper.

\section*{Acknowledgments} 

We thank Chandrasekhar Chatterjee and Simone Giacomelli for useful discussions. 
K.K. thanks the Department of Physics, Osaka City University, for a warm hospitality during his visit there where part of the work was done. 
K.O. and M.N. thank INFN, Pisa, and the Department of Physics, University of Pisa, for a warm hospitality during their visit there where part of the work was done.  J.E. is supported by the Chinese Academy of Sciences
Fellowship for Young International Scientists grant number 2010Y2JA01. 
M.N. is supported in part by Grant-in-Aid for Scientific Research (No. 25400268) 
and by the ``Topological Quantum Phenomena'' 
Grant-in-Aid for Scientific Research 
on Innovative Areas (No. 25103720)  
from the Ministry of Education, Culture, Sports, Science and Technology 
(MEXT) of Japan.  W.V. is supported by the Global Engagement for Global Impact programme funded by EPSRC, and with support from ARO grant number W911NF-12-1-0523.
K.K. is supported by INFN special project, ``Nonperturbative Dynamics in Gauge Theories and in
String Theory (PI14)''.

\appendix

\section{Moduli matrix formalism}

\subsection{BPS equations}
The standard covariant derivative 
has the form, 
\begin{eqnarray}
 {\cal D}_\mu q^A=\partial_\mu q^A+i(A_\mu)^A{}_B q^B,\quad 
(A_\mu)^A{}_B =\sum_I A_\mu^I g_I (T_I)^A{}_B\;.
\end{eqnarray}
The BPS equations for chiral fields
\begin{eqnarray}
 {\cal D}_{\bar z}q^A=0\;,
\end{eqnarray}
where \[z \equiv x+iy\;,  \quad {\bar z}\equiv  x-iy\;, \quad  \partial =\partial_{z}\equiv  \tfrac{1}{2} (\partial_{x}-  i \partial_{y})\;, \quad 
{\bar \partial}\equiv  \tfrac{1}{2} (\partial_{x} + i \partial_{y}) \] 
as usual, 
are solved by    (${\bar A}\equiv  A_{x} + i A_{y}$)
\begin{eqnarray}
 i \bar A= S^{-1}\bar \partial S\;,\quad q^A=(S^{-1})^{A}{}_B
  H_0^B(z)\;,\quad \partial_{\bar z}H_0^A(z)=0\;.
\end{eqnarray}
Introducing a derivative {\it operator}  $\hat {\bar \partial}$, the covariant derivative can be set in a more compact form,
\begin{eqnarray}
  {\cal D}_{\bar z}= {\bar \partial}+i \bar A=S^{-1} \hat {\bar \partial} S\;.
\end{eqnarray}
\begin{eqnarray}
 F_{12}&=&-i[{\cal D}_1,{\cal D}_2]=-2[{\cal D}_z\;, {\cal D}_{\bar z}]=
-2i F_{z\bar z}\nn
&=&-2[S^\dagger \hat \partial S^{\dagger -1}\;,  S^{-1} \hat {\bar
\partial} S]
=-2S^\dagger \left[\, \hat \partial,\, S^{\dagger -1} S^{-1} \hat {\bar
\partial} S S^\dagger \right]  S^{\dagger -1}\nn
&=&-2S^\dagger \left[\, \hat \partial,\, \Omega^{-1} \hat {\bar
\partial} \Omega \right]S^{\dagger -1}=-2S^\dagger 
  \partial  \left(\Omega^{-1} {\bar \partial} \Omega \right)S^{\dagger -1}\;,
\end{eqnarray}
where we used $C [A, B] C^{-1}=[C A C^{-1},C B C^{-1}]$.
BPS equations for gauge fields 
\begin{eqnarray}
 F_{12}^I-g_I \left(q^\dagger_A(T_I q)^A-\xi_I\right)=0
\end{eqnarray}
can be rewritten to
\begin{eqnarray}
 2  \partial  \left(\Omega^{-1} {\bar \partial} \Omega \right)&=&
-S^{\dagger-1}(F_{12}^Ig_I T_I )S^\dagger \nn
&=&g_I^2  S^{\dagger-1} \xi_I T_I S^\dagger
-g_I^2(S^{\dagger-1}T_IS^\dagger) 
\left(H_0^\dagger S^{\dagger -1}T_I S^{-1} H_0\right)\nn
&=&g_I^2\xi_I T_I-g_I^2T_I \left(H_0^\dagger T_I \Omega^{-1}H_0\right)
\end{eqnarray}
where we assume that,  under complexified gauge transformation with
$C\in G^{\mathbb C}$, $\zeta$ is invariant as 
\begin{eqnarray}
 C \zeta C^{-1}= \zeta\;,\qquad  \zeta \equiv \sum_Ig_I^2\xi_I T_I\;,
\end{eqnarray}
and a tensor $\sum_I T_I \otimes T_I$ is also invariant
\begin{eqnarray}
 \sum_{I} C T_I C^{-1}\otimes C T_I C^{-1}=\sum_I T_I \otimes T_I
\end{eqnarray}
Therefore we find that  the master equation 
\begin{eqnarray}
 2  \partial  \left(\Omega^{-1} {\bar \partial} \Omega \right)=
\zeta-g_I^2T_I \left(H_0^\dagger T_I \Omega^{-1}H_0\right)\;.
\end{eqnarray}
 is  equivalent to
\begin{eqnarray}
2  \bar \partial  \left({\partial} \Omega\, \Omega^{-1}\right)=
\zeta-g_I^2T_I \left(H_0^\dagger\Omega^{-1} T_I H_0\right)\;, 
\end{eqnarray}
where we used the following identity 
\begin{eqnarray}
 \Omega \partial  \left(\Omega^{-1} {\bar \partial} \Omega \right)
  \Omega^{-1}=\bar \partial \left({\partial} \Omega\, \Omega^{-1}\right)\;.
\end{eqnarray}

\subsection{Zero modes and Gauss's law}

Following Ref.~\cite{Eto:2006uw}, 
let us now lift the moduli parameters in the moduli matrix $H_0^A$ to chiral fields 
\begin{eqnarray}
 H_0^A(z,\phi^X) \to H_0\left(z,\phi^X(x^\alpha)\right)\;,  \qquad x^{\alpha}= x^{3}, x^{0}\;.
\end{eqnarray}
By assumption  $H_0$ contains no anti-chiral field so that
\begin{eqnarray}
 \delta_\alpha ^\dagger H_0^A=0
\end{eqnarray}
where $\delta_\alpha, \delta_\alpha^\dagger $ are defined by
\begin{eqnarray}
 \delta_\alpha =\partial_\alpha \phi^X \frac{\delta}{\delta
  \phi^X}\;,\quad
\delta_\alpha^\dagger  =\partial_\alpha \bar \phi^X \frac{\delta}{\delta
  \bar \phi^X}\;.
\end{eqnarray}
We can show that with the following gauge fields parallel to  the vortex
configuration ($\alpha=0,3$),  
\begin{eqnarray}
 i A_\alpha=S^{-1}( \delta_\alpha^\dagger  S)+S^\dagger 
  (\delta_\alpha S^{\dagger-1})\;,\quad {\cal D}_\alpha=S^{-1}\hat \delta_\alpha^\dagger  S+S^\dagger \hat
  \delta_\alpha S^{\dagger-1} 
\end{eqnarray}
satisfy the Gauss law. 
First we find  
\begin{eqnarray}
 {\cal D}_\alpha q=S^\dagger \delta_\alpha\left(\Omega^{-1}H_0\right)
\end{eqnarray}
and 
\begin{eqnarray}
 iF_{\alpha \bar z}&=&[{\cal D}_\alpha,\,{\cal D}_{\bar z}]=
[S^{-1}\hat \delta_\alpha^\dagger  S,\, S^{-1}\hat {\bar \partial} S]+
[S^\dagger \hat
  \delta_\alpha S^{\dagger-1},\, S^{-1}\hat {\bar \partial} S]\nn
&=&S^{-1}[\hat \delta_\alpha^\dagger,\, \hat {\bar \partial}] S+
S^\dagger [\hat
  \delta_\alpha,\, \Omega^{-1}\hat {\bar \partial} \Omega
  ]S^{\dagger-1}\nn
&=&S^\dagger \delta_\alpha
\left(\Omega^{-1}{\bar \partial} \Omega\right)S^{\dagger-1}.
\end{eqnarray}
By using these, next, we find  
\begin{eqnarray}
\left(q^\dagger g_IT_I {\cal D}_\alpha q \right)\otimes g_I T_I &=&
\left(H_0^\dagger g_I^2T_I \delta_\alpha (\Omega^{-1}H_0)\right)\otimes
(S^{\dagger}T_I S^{\dagger-1})\nn
&=&-2 S^\dagger \delta_\alpha \left(\partial (\Omega^{-1}\bar \partial
			       \Omega)\right) S^{\dagger-1}
\end{eqnarray}
and
\begin{eqnarray}
 2i{\cal D}F_{\alpha \bar z}^I\, (g_I T_I)&=&2i{\cal D}F_{\alpha \bar z}\nn
&=&2 \left[S^\dagger \hat \partial S^{\dagger-1}, S^\dagger \delta_\alpha
\left(\Omega^{-1}{\bar \partial} \Omega\right)S^{\dagger-1}\right]\nn
&=&2 S^\dagger \left[ \hat \partial ,  \delta_\alpha
\left(\Omega^{-1}{\bar \partial} \Omega\right) \right] S^{\dagger-1}
=2 S^\dagger   \partial   \delta_\alpha
\left(\Omega^{-1}{\bar \partial} \Omega\right) S^{\dagger-1}.\nn
\end{eqnarray}
This shows that (half of) Gauss's law 
\begin{eqnarray}
 q^\dagger g_IT_I {\cal D}_\alpha q +2i{\cal D}F_{\alpha \bar z}^I=0
\end{eqnarray}
is indeed satisfied. 

\subsection{The low-energy effective action   \label{effeAction}}

Next we calculate the effective action,
\begin{eqnarray}
 {\cal L}_{\rm eff}=\int d^2x \left(
 \frac12 (F_{\alpha i}^I )^2
+{\cal D}_\alpha q^\dagger {\cal D}^\alpha q\right).
\end{eqnarray}
Two terms on the right hand side are given by
\begin{eqnarray}
 {\cal D}_\alpha q^\dagger {\cal D}^\alpha q &=&
\delta_\alpha^\dagger \left(H_0^\dagger \Omega^{-1}\right)\Omega \, 
\delta_\alpha\left(\Omega^{-1}H_0\right)\nn
&=&\delta_\alpha^\dagger \left(H_0^\dagger \Omega^{-1}\right)
\delta_\alpha H_0 -\sum_I \delta_\alpha^\dagger \left(H_0^\dagger \Omega^{-1}\right)
g_IT_I H_0 \,  (\delta_\alpha \Omega\,\Omega^{-1})^I\nn
&=&\delta_\alpha^\dagger \left(H_0^\dagger \Omega^{-1}\delta_\alpha
			  H_0\right)
-\sum_I\delta_\alpha^\dagger \left(H_0^\dagger \Omega^{-1}
T_I H_0 \right) \,  g_I(\delta_\alpha \Omega\,\Omega^{-1})^I  \;,  \label{scalar}
\end{eqnarray}
and
\begin{eqnarray}
 \frac12 (F_{\alpha i}^I )^2&=& 2 |F_{\alpha \bar z}^I|^2
=\sum_{I}\frac{2}{g_I^2c^2} 
\left|{\rm Tr}\left[T_I S^\dagger \delta_\alpha
\left(\Omega^{-1}{\bar \partial} \Omega\right)S^{\dagger-1}\right]
\right|^2\nn
&=&\sum_{I}\frac{2}{g_I^2c^2} 
{\rm Tr}\left[T_I S^\dagger \delta_\alpha
\left(\Omega^{-1}{\bar \partial} \Omega\right)S^{\dagger-1}\right]
{\rm Tr}\left[T_I S^{-1} \delta_\alpha^\dagger
\left({\partial} \Omega\, \Omega^{-1}\right)S\right]\nn
&=&\sum_{I}\frac{2}{g_I^2c^2} 
{\rm Tr}\left[T_I \Omega \delta_\alpha
\left(\Omega^{-1}{\bar \partial} \Omega\right)\Omega^{-1}\right]
{\rm Tr}\left[T_I \delta_\alpha^\dagger
\left({\partial} \Omega\, \Omega^{-1}\right)\right]\nn
&=&\sum_{I}\frac{2}{g_I^2c^2} 
{\rm Tr}\left[T_I\bar \partial\left(\delta_\alpha \Omega
			       \Omega^{-1}\right)\right]
{\rm Tr}\left[T_I \delta_\alpha^\dagger
\left({\partial} \Omega\, \Omega^{-1}\right)\right]\nn
&=&\bar \partial \left\{ \Sigma_{z}  \right\}
+\Delta {\cal L}\;,      \label{gauge}
\end{eqnarray}
where
\begin{eqnarray}
 \Sigma_{z}&=&\sum_{I}\frac{2}{g_I^2c^2} 
{\rm Tr}\left[T_I \delta_\alpha \Omega
			       \Omega^{-1} \right]
{\rm Tr}\left[T_I \delta_\alpha^\dagger
\left({\partial} \Omega\, \Omega^{-1}\right)\right]\nn
&=&\sum_{I}\frac{2}{g_I^2c^2} 
{\rm Tr}\left[T_I \delta_\alpha \Omega
			       \Omega^{-1} \right]
{\rm Tr}\left[T_I \Omega \partial 
\left(\Omega^{-1}\,\delta_\alpha^\dagger \Omega\right)\Omega^{-1}\right]\nn
&=&\sum_{I}\frac{2}{g_I^2c^2} 
{\rm Tr}\left[T_I\Omega^{-1}\delta_\alpha \Omega
			        \right]
\partial 
\left({\rm Tr}\left[T_I \Omega^{-1}\,\delta_\alpha^\dagger \Omega
\right]\right)
\end{eqnarray}
and 
\begin{eqnarray}
\Delta {\cal L}&=&-\sum_{I}\frac{2}{g_I^2c^2} 
{\rm Tr}\left[T_I \delta_\alpha \Omega
			       \Omega^{-1}\right]
{\rm Tr}\left[T_I \delta_\alpha^\dagger \bar \partial
\left({\partial} \Omega\, \Omega^{-1}\right)\right] \nn
&=&\sum_{I}\frac{1}{c^2} 
{\rm Tr}\left[T_I \delta_\alpha \Omega
			       \Omega^{-1}\right]
{\rm Tr}[T_IT_J]
\delta_\alpha^\dagger \left(H_0^\dagger \Omega^{-1}
T_J H_0 \right)\nn
&=&\sum_{I}
g_I(\delta_\alpha \Omega \Omega^{-1})^I
\delta_\alpha^\dagger \left(H_0^\dagger \Omega^{-1}
T_I H_0 \right)\;.
\end{eqnarray}
Here the following normalization of generator of $G$ are used:
\begin{eqnarray}
 {\rm Tr}[T_I T_J]= c\, \delta_{IJ}
\end{eqnarray}
and  $X=X^I(g_I T_I),\quad Y=Y^I(g_I T_I)$,  
\begin{eqnarray}
 X^I= \frac1{g_I c}{\rm Tr}[T_I X]\;,\quad \sum_{I}{\rm Tr}[T_I X]{\rm
  Tr}[T_I Y]=c \,{\rm Tr}[X Y]\;. 
\end{eqnarray}
We see that $\Delta {\cal L}$ cancels between the gauge and scalar kinetic terms, and the effective Lagrangian can be written in the form, 
\begin{eqnarray}
 {\cal L}_{\rm eff}=\frac{\partial^2 K_{\rm core}}{\partial \phi^X
  \partial \bar \phi^Y } \partial_\alpha \phi^X \partial^\alpha \bar
  \phi^Y+\frac{\partial^2 K_{\rm bulk}}{\partial \phi^X
  \partial \bar \phi^Y } \partial_\alpha \phi^X \partial^\alpha \bar \phi^Y\;, 
\end{eqnarray}
where
\begin{eqnarray}
 \frac{\partial K_{\rm core}}{\partial \phi^X }=\int d^2x H_0^\dagger
  \Omega^{-1}\frac{\partial H_0}{\partial \phi^X}
\end{eqnarray}
and 
\begin{eqnarray}
 \frac{\partial^2 K_{\rm bulk}}{\partial \phi^X
  \partial \bar \phi^Y } \partial_\alpha \phi^X \partial^\alpha \bar
  \phi^Y
=-\frac{i}2 \int_{|z|=\Lambda} d z \Sigma_z\;,
\end{eqnarray}
see Eqs.~(\ref{scalar}) and (\ref{gauge}).   

Applied to our concrete $SU(2)\times U(1)$ model, these give 
(read  $\phi^{X}  \to B;$   $\bar \phi^Y   \to {\bar B;}$)
\beq       K = K_{core}+ K_{bulk}\;,
\eeq 
where 
\beq  K_{core} \simeq  K_{core}|_{g_{R}=0}  =  K^{(CP)} =  \frac{4\pi}{{g_{L}^2}}   \log  (|B|^{2}+1)\;,
\eeq
while $K_{bulk}$ is divergent:
\beq    \frac{\partial^{2}  K_{bulk}}{\partial B \partial {\bar B}}   =\frac{4\pi}{e^2}\partial_{\bar B}\beta \,
\partial_B \psi_{em}\Big|_{|z|=\Lambda} =   \frac{16\pi   e^{2}}{g_{L}^{4}}     \frac { |B|^{2}}{(|B|^{2}+1)^{2}} \, \log \Lambda \;.
\eeq

\section{Solutions without $U(1)_{R}$ weak gauging  \label{solutions}}

Let us consider a solution for  $\psi_0,\psi_3,w$  where
\begin{eqnarray}
 S=e^{\psi_0}\left(\begin{array}{cc}
	e^{\psi_3/2}& 0 \\  e^{\psi_3/2} w & e^{-\psi_3/2}
			   \end{array}\right),\qquad 
H_0=\sqrt{\xi}\left(\begin{array}{cc}
     z & 0 \\ -B &1
	  \end{array}\right).
\end{eqnarray}
We call the  solution for $B=0$ as
$\psi_0=\psi_0^{(0)},\psi_3=\psi_3^{(0)},w=0$:
\begin{eqnarray}
 S^{(0)}=e^{\psi_0^{(0)}}\left(\begin{array}{cc}
	e^{\psi_3^{(0)}/2}& 0 \\ 0 & e^{-\psi_3^{(0)}/2}
			   \end{array}\right),\qquad 
H_0^{(0)}=\sqrt{\xi}\left(\begin{array}{cc}
     z & 0 \\ 0 &1
	  \end{array}\right).
\end{eqnarray}
$\psi_0^{(0)}$ and $\psi_3^{(0)}$ are determined explicitly in the main text (Eq.~(\ref{B=0sol}), Eq.~(\ref{Taube})). 
The generic $B\ne 0$ solutions are generated by a color-flavor rotation
\begin{eqnarray}
 V_BH_0=H_0^{(0)}U_B, \qquad U_B=\frac1{\sqrt{1+|B|^2}}\left(\begin{array}{cc}
	1&  B^{*} \\ -B & 1
			   \end{array}\right)\;,
\end{eqnarray}  
where the V-transformation $V_B$ can be taken as
\begin{eqnarray}
 V_B=\frac1{\sqrt{1+|B|^2}}\left(\begin{array}{cc}
	1+|B|^2& B^{*}  z \\ 0 & 1
			   \end{array}\right)
\end{eqnarray} 
so as to bring back $S$ to the lower triangular  form.  
Therefore there must be the following relation with an appropriate
$U(2)_L$ gauge transformation $U_L$  
\begin{eqnarray}
 V_B S=S^{(0)}U_L\;,
\end{eqnarray}
or 
\begin{eqnarray}
 SS^\dagger = V_B^{-1}S^{(0)}S^{(0)\dagger} (V_B^\dagger)^{-1}\;,
\end{eqnarray}
that is
\begin{eqnarray}
e^{\psi_0}\left(\begin{array}{cc}
	e^{\psi_3}& 
e^{\psi_3} \bar w  \\ e^{\psi_3} w & |w|^2e^{\psi_3}+e^{-\psi_3} 
			   \end{array}\right) 
=e^{\psi^{(0)}_0}\left(\begin{array}{cc}
	\frac{e^{\psi^{(0)}_3}+|B|^2|z|^2e^{-\psi^{(0)}_3}}{1+|B|^2}& 
-z \bar B e^{-\psi^{(0)}_3} \\ -\bar z B e^{-\psi^{(0)}_3}  & 
(1+|B|^2)e^{-\psi^{(0)}_3}  			   \end{array}\right)\;.
\end{eqnarray}
Comparing the both sides  we find: 
\begin{eqnarray}
 \psi_0=\psi^{(0)}_0,\qquad
\psi_3=\log\left(
\frac{e^{\psi^{(0)}_3}+|B|^2|z|^2e^{-\psi^{(0)}_3}}{1+|B|^2}\right)\;,  \nn
w=-\bar z B e^{-\psi_3-\psi_3^{(0)}}
=\frac{-\bar z B (1+|B|^2)e^{-\psi_3^{(0)}}}
{e^{\psi^{(0)}_3}+|B|^2|z|^2e^{-\psi^{(0)}_3}}\;.
\end{eqnarray}

\end{document}